\documentclass[twocolumn,showpacs,showkeys,preprintnumbers,amssymb,aps,superscriptaddress,prb]{revtex4}
\usepackage{graphicx}
\usepackage{dcolumn}
\usepackage{color}
\usepackage{bm}
\usepackage{dsfont}
\usepackage{enumitem}
\usepackage{epstopdf}

\begin{document}


\title{Cluster-based Haldane phases, bound magnon crystals and quantum spin liquids of a mixed spin-1 and spin-1/2 Heisenberg octahedral chain}
\author{Katar\'ina Karl'ov\'a}
\affiliation{Institute of Physics, Faculty of Science, P. J. \v{S}af\'{a}rik University, Park Angelinum 9, 04001 Ko\v{s}ice, Slovakia}
\author{Jozef Stre\v{c}ka}
\email{jozef.strecka@upjs.sk}
\affiliation{Institute of Physics, Faculty of Science, P. J. \v{S}af\'{a}rik University, Park Angelinum 9, 04001 Ko\v{s}ice, Slovakia}
\author{Taras Verkholyak}
\affiliation{Institute for Condensed Matter Physics, NASU, Svientsitskii Street 1, 79011 L'viv, Ukraine}

\date{\today}

\begin{abstract}
The mixed spin-1 and spin-1/2 Heisenberg octahedral chain with regularly alternating monomeric spin-1 sites and square-plaquette spin-1/2 sites is investigated using variational technique, localized-magnon approach, exact diagonalization ({\ ED}) and density-matrix renormalization group ({\ DMRG}) method. The investigated model has in a magnetic field an extraordinarily rich ground-state phase diagram, which includes the uniform and cluster-based Haldane phases, two ferrimagnetic phases of Lieb-Mattis type, two quantum spin liquids and two bound magnon crystals in addition to the fully polarized ferromagnetic phase. The lowest-energy eigenstates in a highly-frustrated parameter region belong to flat bands and hence, low-temperature thermodynamics above the bound magnon-crystal ground states can be satisfactorily described within the localized-magnon approach. The variational method provides an exact evidence for the magnon-crystal phase with a character of the monomer-tetramer ground state at zero field, while another magnon-crystal phase with a single bound magnon at each square plaquette is found in a high-field region. A diversity of quantum ground states gives rise to manifold zero-temperature magnetization curves, which may involve up to four wide intermediate plateaus at zero, one-sixth, one-third and two-thirds of the saturation magnetization, two quantum spin-liquid regions and two tiny plateaus at one-ninth and one-twelfth of the saturation magnetization corresponding to the fragmentized cluster-based Haldane phases.

\end{abstract}

\maketitle

\section{Introduction}
Quantum Heisenberg systems with antiferromagnetic interactions may exhibit unconventional quantum ground states, which are responsible for anomalous course of magnetization curves at low temperatures. One of the most spectacular quantum phases with an energy gap is topologically  nontrivial Haldane-type phase, which has been discovered in the ground state of the antiferromagnetic spin-1  Heisenberg chain.\cite{hald83,haldan83} A subtle nature of the Haldane phase has been revealed by Affleck, Kennedy, Lieb, and Tasaki when considering a more general bilinear-biquadratic version of the antiferromagnetic spin-1 Heisenberg chain (the so-called AKLT model), which has an exact valence-bond-solid ground state with only slightly higher (about 5\%) ground-state energy in comparison with the Haldane phase.\cite{affl87,affl88}  It is noteworthy that the valence-bond-solid ground state of the AKLT model has the maximal value of the hidden string-order parameter, which is ultimately connected with creating singlets between all adjacent spin-1 particles symmetrically decomposed into two fictitious spin-1/2 quasiparticles. Owing to this fact, the valence-bond-solid ground state of the AKLT model has a character of a unique singlet state with topologically protected spin-1/2 edges and the same holds true for the Haldane phase being continuously connected to the valence-bond-solid ground state of the AKLT model.\cite{LNP2004} {\ Electron-spin-resonance measurements on the nickel-based compound NENP being an experimental realization of the antiferromagnetic spin-1 Heisenberg chain are consistent with breaking of valence bonds (singlets) achieved through a doping with the spin-1/2 impurities,\cite{hagi90} whereas inelastic-neutron-scattering data for pure and doped samples of another nickel-based compound Y$_2$BaNiO$_5$ afforded a more direct evidence of edge states.\cite{kenz03}} These experimental observations have thus verified proximity of the Haldane phase to the valence-bond-solid ground state.

From the viewpoint of the magnetization response, the Haldane phase macroscopically manifests itself in a zero-temperature magnetization curve as a zero plateau, which breaks down at a critical field closing a singlet-triplet energy gap associated with a field-driven quantum phase transition towards the Tomonaga-Luttinger quantum spin liquid.\cite{misguitch} It is worthwhile to remark that intermediate magnetization plateaus emergent in magnetization process of quantum Heisenberg antiferromagnets may correspond to exotic quantum phases of diverse character and they can occur at different fractional values of the saturation magnetization.\cite{rich04,taki,taki2} Oshikawa, Yamanaka, and Affleck have found a rigorous criterion for an existence of the intermediate magnetization plateaus in one-dimensional quantum Heisenberg systems, which may exhibit magnetization plateaus on assumption that the following quantization condition is met: $p(S_u - m_u) \in \mathbb{Z}$, where $p$ is a period of the ground state,  $S_u$ denotes the total spin of elementary unit,  $m_u$ determines the total magnetization of elementary unit and  $ \mathbb{Z}$ is a set of integer numbers.\cite{oshi97,affl98} 

The aforementioned quantization condition would imply that the higher the period of the ground state is, the greater is the number of available magnetization plateaus. The spin-1/2 Heisenberg orthogonal-dimer chain \cite{schu02,schul02,johannes,verk16} represents a rare example of a quantum spin chain, which exhibits in a zero-temperature magnetization curve a peculiar sequence of infinite number of intermediate magnetization plateaus. By numerical exact diagonalization  Schulenburg and Richter  \cite{schul02} have rigorously proved  that  infinite series of magnetization plateaus of the spin-1/2   Heisenberg orthogonal-dimer chain at $\frac{\mathbb{Z}}{2(\mathbb{Z}+1)}$ of the saturation magnetization is a direct consequence of the fragmentation of the magnetic ground state.

The fragmentation, which is caused by a local creation of singlets, may be also responsible for existence of cluster-based Haldane-type phases as originally reported for the Heisenberg diamond chain.\cite{taka96,hida10,hida11,hida17} Recent experimental discovery {\ of} cluster-based Haldane-type phases in minerals fedotovite K$_2$Cu$_3$O(SO$_4$)$_3$,\cite{fuji18} euchlorine KNaCu$_3$O(SO$_4$)$_3$ and puninite Na$_2$Cu$_3$O(SO$_4$)$_3$\cite{furr18} has stimulated a renewed interest in a search of other quantum spin chains possibly displaying the cluster-based Haldane-type ground states.\cite{stre18,sugiarx}
Another exotic quantum ground states, which are manifested in a zero-temperature magnetization curve as intermediate magnetization plateaus emergent just below the saturation field, may have a character of the bound magnon crystals. It is worth mentioning that the localized nature of bound magnons within the magnon-crystal phases enables a description of low-temperature magnetization curves from a mapping correspondence with  classical lattice-gas (Ising) models.\cite{zhit05,derz06,derz15}  In our recent work, we have found that the localized-magnon approach can be extended to cover a full magnetization curve of the spin-1/2 Heisenberg octahedral chain from zero up to saturation field whenever the lowest-energy bound one- and two-magnon states are simultaneously taken into consideration.\cite{stre17,strepb}

 The present work will be devoted to a detailed examination of the mixed spin-1 and spin-1/2 Heisenberg octahedral chain in a magnetic field. It will be demonstrated hereafter that the considered quantum spin chain exhibits a lot of intriguing quantum ground states including the cluster-based Haldane phases, the bound magnon-crystal phases, the Lieb-Mattis ferrimagnetic phases and the Tomonaga-Luttinger quantum spin liquids, which will be the main subject of our investigations. 

This paper is organized as follows. The mixed-spin Heisenberg octahedral chain will be defined in Sec. \ref{sec:model}, in which several complementary calculation techniques will also be presented. The most interesting results for the ground-state phase diagram, magnetization curves and specific heat are discussed in Sec. \ref{sec:result}. Finally, several concluding remarks are mentioned in Sec. \ref{sec:conc}.

\section{Heisenberg octahedral chain}
\label{sec:model}

\begin{figure}
\begin{center}
\includegraphics[width=0.45\textwidth]{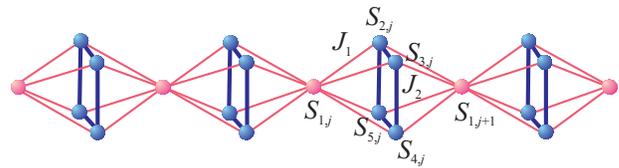}
\end{center}
\vspace{-0.6cm}
\caption{(Color online) A part of the mixed spin-1 {\ (light red spheres)} and spin-1/2 {\ (dark blue spheres)} Heisenberg octahedral chain. Thick (blue) lines denote the Heisenberg intra-plaquette interaction $J_2$, while thin (red) lines represent the monomer-plaquette coupling $J_1$.}
\label{fig1}
\end{figure}

In the present work, we will explore the mixed spin-1 and spin-1/2 Heisenberg octahedral chain diagrammatically illustrated in Fig.~\ref{fig1}, which can be defined through the following Hamiltonian
\begin{eqnarray}
\label{ham}
\hat{\cal H} \!\!&=&\!\! 
\sum_{j=1}^{N} \Bigl[ J_1 (\boldsymbol{\hat{S}}_{1,j} + \boldsymbol{\hat{S}}_{1,j+1}) \!\cdot\! (\boldsymbol{\hat{S}}_{2,j} + \boldsymbol{\hat{S}}_{3,j} + \boldsymbol{\hat{S}}_{4,j} + \boldsymbol{\hat{S}}_{5,j}) \Bigr.  \nonumber \\
\!\!&+&\!\! J_2 (\boldsymbol{\hat{S}}_{2,j}\!\cdot\!\boldsymbol{\hat{S}}_{3,j} + \boldsymbol{\hat{S}}_{3,j}\!\cdot\!\boldsymbol{\hat{S}}_{4,j}
+ \boldsymbol{\hat{S}}_{4,j}\!\cdot\!\boldsymbol{\hat{S}}_{5,j} + \boldsymbol{\hat{S}}_{5,j}\!\cdot\!\boldsymbol{\hat{S}}_{2,j}) \nonumber \\
\Bigl. \!\!&-&\!\! h \sum_{i=1}^{5} \hat{S}_{i,j}^{z} \Bigr].
\end{eqnarray}
Here, $\boldsymbol{\hat{S}}_{i,j} \equiv (\hat{S}_{i,j}^x, \hat{S}_{i,j}^y, \hat{S}_{i,j}^z)$ labels three spatial components of the spin-1 (spin-1/2) operator for the unit cell index $i=1$ ($i=2,3,4,5 $), respectively. The exchange interaction $J_1>0$ accounts for the antiferromagnetic monomer-plaquette interaction between the nearest-neighbor spin-1 and spin-1/2 particles, while the interaction constant {\ $J_2>0$} accounts for the antiferromagnetic intra-plaquette interaction between the nearest-neighbor spins-1/2 particles from the same square plaquette. The last term in the Hamiltonian (\ref{ham}) {\ represents the standard Zeeman's term for magnetic moments in an external magnetic field $h \geq 0$}. The periodic boundary condition $\boldsymbol{S}_{1,N+1} \equiv \boldsymbol{S}_{1,1}$ is assumed  for simplicity. Let us examine the Hamiltonian (\ref{ham}) by employing a few complementary analytical and numerical techniques thoroughly described in the subsequent subsections.

\subsection{Variational principle}
\label{vm} 

\begin{figure}
\begin{center}
\includegraphics[width=0.15\textwidth]{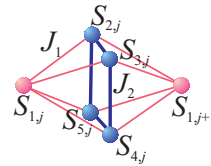}
\end{center}
\vspace{-0.6cm}
\caption{(Color online) A six-spin cluster with the geometric shape of an elementary octahedron, which is used to built up the mixed spin-1 and spin-1/2 Heisenberg octahedral chain.}
\label{fig2}
\end{figure}

In a highly frustrated parameter region one may adapt a variational technique\cite{shas81,bose89,bose90,bose92} in order to find an exact ground state of the mixed spin-1 and spin-1/2 Heisenberg octahedral chain in a zero magnetic field. The main idea of this approach lies in decomposing the total Hamiltonian (\ref{ham}) into a sum of cell Hamiltonians $\hat{\cal H} = \sum_{j=1}^N\hat{\cal H}_{j}$, where each cell Hamiltonian $\hat{\cal H}_{j}$ pertinent to a six-spin cluster with a geometric shape of an octahedron (see Fig. \ref{fig2}) is given by
\begin{eqnarray}
\hat{\cal H}_{j} \!\!&=&\!\! J_1 \left(\boldsymbol{\hat{S}}_{1,j}+ \boldsymbol{\hat{S}}_{1,j+1}\right) \cdot (\boldsymbol{\hat{S}}_{2,j} + \boldsymbol{\hat{S}}_{3,j} 
                                 + \boldsymbol{\hat{S}}_{4,j} + \boldsymbol{\hat{S}}_{5,j}) \nonumber \\
                  \!\!&+&\!\! J_2 (\boldsymbol{\hat{S}}_{2,j}\!\cdot\!\boldsymbol{\hat{S}}_{3,j} + \boldsymbol{\hat{S}}_{3,j}\!\cdot\!\boldsymbol{\hat{S}}_{4,j}
+ \boldsymbol{\hat{S}}_{4,j}\!\cdot\!\boldsymbol{\hat{S}}_{5,j} + \boldsymbol{\hat{S}}_{5,j}\!\cdot\!\boldsymbol{\hat{S}}_{2,j}). \nonumber \\
\label{hamclu}
\end{eqnarray}
Subsequently, the variational arguments can be used term-by-term in order to obtain a lower bound for the ground-state energy of the mixed spin-1 and spin-1/2 Heisenberg octahedral chain $E_0$ from a sum of the lowest eigenenergies $\varepsilon_{j}^0$ of cell Hamiltonians, i.e. $E_0\geq \sum_{j=1}^{N}\varepsilon_{j}^0$. The energy spectrum of a single mixed-spin Heisenberg octahedron with apical spin-1 particles and a square base composed of four spin-1/2 particles (see Fig. \ref{fig2} for illustration) can be expressed in terms of five quantum spin numbers $S_{T,j}$, $S_{\square,j}$, $S_{24,j}$, $S_{35,j}$ and $S_{16,j}$
\begin{eqnarray}
\varepsilon_{j} \!\!&=&\!\! \frac{J_1}{2} S_{T,j}(S_{T,j}+1) + \frac{J_2-J_1}{2} S_{\square,j} (S_{\square,j} + 1) \nonumber \\
    \!\!&-&\!\! \frac{J_2}{2} [S_{24,j} (S_{24,j} + 1) + S_{35,j} (S_{35,j} + 1)] \nonumber \\
    \!\!&-&\!\!  \frac{J_1}{2}S_{16,j}(S_{16,j} + 1),
\label{spec}
\end{eqnarray}
which determine the total spin of the octahedron $S_{T,j}$, the total spin of the square plaquette $S_{\square,j}$, the total spin of the spin-1 apical particles $S_{16,j}$ and the total spin of two spin-1/2 particles from opposite corners of a square plaquette $S_{24,j}$ and $S_{35,j}$, respectively. It follows from Eq. (\ref{spec}) that the lowest-energy eigenstate of the mixed-spin Heisenberg octahedron in a highly frustrated parameter region $J_2>3J_1$ is a degenerate state characterized by the quantum spin numbers $S_{\square,j} = 0$, $S_{24,j}= 1$, $S_{35,j} = 1$ and $S_{T,j,k} = S_{16,j}$, whereas the latter quantum spin numbers $S_{T,j,k}$ and $S_{16,j}$ can take any out of three possible values 0, 1, 2. This result is taken to mean that the four spins-1/2 particles from each square base are in a singlet-tetramer state
and the spin-1 particles from the apical (monomeric) sites become completely free, i.e. paramagnetic in character because the singlet-tetramer state breaks correlation between the monomeric spins on both its sides. The monomer-tetramer (MT) phase thus becomes the relevant ground state of the mixed spin-1 and spin-1/2 Heisenberg octahedral chain with the following eigenvector
\begin{eqnarray}
|{\rm MT} \rangle \!\!=\!\! \prod_{j=1}^N \!\! |\rm{0,\pm 1}\rangle_{1,j} \!\!\otimes\!\! 
\Bigl[\!\frac{1}{\sqrt{3}}(|\!\!\uparrow_{\!2,j}\downarrow_{3,j}\uparrow_{\!4,j}\downarrow_{5,j}\rangle \!\!\!\!&+&\!\!\!\! |\!\!\downarrow_{2,j}\uparrow_{\!3,j}\downarrow_{4,j}\uparrow_{\!5,j}\!\rangle\!)  \nonumber \\
- \frac{1}{\sqrt{12}} (|\!\!\uparrow_{2,j}\uparrow_{3,j}\downarrow_{4,j}\downarrow_{5,j}\rangle \!\!\!\!&+&\!\!\!\! |\!\!\uparrow_{2,j}\downarrow_{3,j}\downarrow_{4,j}\uparrow_{5,j}\rangle \nonumber \\
 \hspace{-0.7cm}\!\!\!\!\!\!\!\!\!\!+\! |\!\!\downarrow_{2,j}\uparrow_{3,j}\uparrow_{4,j}\downarrow_{5,j}\rangle \!\!\!\!&+&\!\!\!\! |\!\!\downarrow_{2,j}\downarrow_{3,j}\!\uparrow_{4,j}\!\uparrow_{5,j}\rangle\!)\! \Bigr]\!, \nonumber \\
\label{MT}
\end{eqnarray}
where the former eigenvectors refer to the monomeric spin-1 particles and the latter eigenvector specifies state of the plaquette spin-1/2 particles.

The variational principle consequently proves emergence of the exact MT ground state provided that the condition $J_2>3J_1$ is met. Under this circumstance the apical spin-1 particles are completely free in a zero field and become fully polarized by any nonzero external magnetic field due to a nonmagnetic character of the singlet-plaquette state of four spin-1/2 particles forming a square base. Owing to this fact, the MT ground state should manifest itself in a zero-temperature magnetization curve as the intermediate one-third plateau with regard to a full polarization of the apical spin-1 particles. Note furthermore that the singlet-tetramer state is nothing but the localized two-magnon state. 
 
\subsection{Localized-magnon approach}
\label{lmgs}

\begin{figure}
\begin{center}
\includegraphics[width=0.5\textwidth]{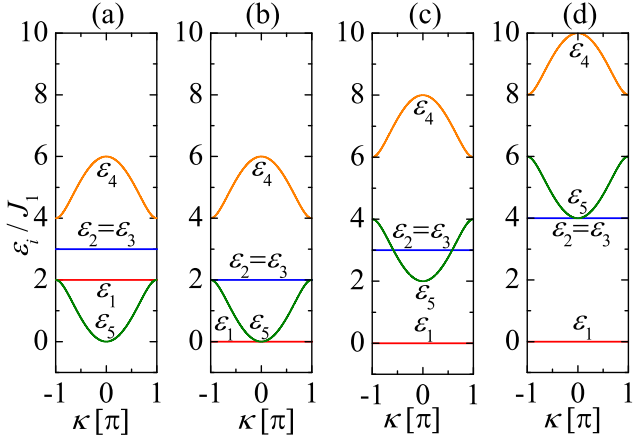}
\end{center}
\vspace{-0.8cm}
\caption{{(Color online) The one-magnon energy bands (\ref{oms}) of the mixed spin-1 and spin-1/2 Heisenberg octahedral chain at the saturation field for four different values of the interaction ratio: (a) $J_2/J_1 = 1$, $h_s/J_1 = 6$; (b) $J_2/J_1 = 2$, $h_s/J_1 = 6$; (c) $J_2/J_1 = 3$, $h_s/J_1 = 8$; (d) $J_2/J_1 = 4$, $h_s/J_1 = 10$.}}
\label{fig3}
\end{figure}

The fully polarized ferromagnetic (FM) state represents another ground state of the mixed spin-1 and spin-1/2 Heisenberg octahedral chain 
\begin{eqnarray}
|{\rm FM} \rangle = \prod_{j=1}^N \!  |\rm{1}\rangle_{1,j}\otimes|\uparrow_{2,j}\uparrow_{3,j}\uparrow_{4,j}\uparrow_{5,j}\rangle, 
\label{FM}
\end{eqnarray}
which emerges at high enough magnetic fields and has the following energy $E_{\rm FM} = N (4J_1 + J_2) - 3N h$. In the frustrated parameter space one may adapt the concept of independent localized magnons \cite{derz06,derz15} in order to determine an exact ground state emergent below the saturation field. The exact one-magnon eigenstates can be found with the help of orthonormal basis $|i, j \rangle = \hat{S}_{i,j}^{-} |{\rm FM} \rangle$ ($i=1-5$, $j=1-N$), which forms the sector $S_T^z = 3N - 1$ with a single spin deviation from the fully polarized FM state. If the Hamiltonian (\ref{ham}) is applied on a given basis set one gets the following set of equations
{\
\begin{eqnarray}
\hat{\cal H} |1, j\rangle \!\!\!&=&\!\!\! (E_{\rm FM} \!+\! h \!-\! 4 J_1) |1, j\rangle  + \frac{J_1}{\sqrt{2}} \sum_{i=2}^5 (|i, j\!-\!1\rangle \!+\! |i, j\rangle), \nonumber \\
\hat{\cal H} |k, j\rangle \!\!\!&=&\!\!\! (E_{\rm FM} + h - 2J_1 - J_2) |k, j\rangle  + \frac{J_2}{2} (|3, j\rangle + |5, j\rangle) \nonumber \\
\!\!\!&+&\!\!\! \frac{J_1}{\sqrt{2}} (|1, j\rangle + |1, j+1\rangle)  \qquad (\mbox{for} \, k = 2 \,\, \mbox{or} \,\, 4), \nonumber \\
\hat{\cal H} |l, j\rangle \!\!\!&=&\!\!\! (E_{\rm FM} + h - 2J_1 - J_2) |l, j\rangle  + \frac{J_2}{2} (|2, j\rangle + |4, j\rangle) \nonumber \\
\!\!\!&+&\!\!\! \frac{J_1}{\sqrt{2}} (|1, j\rangle + |1, j+1\rangle) \qquad (\mbox{for} \, l = 3 \,\, \mbox{or} \,\, 5). 
\label{lm}
\end{eqnarray}}
The set of equations (\ref{lm}) can be subsequently used for solving the eigenvalue problem in  the one-magnon sector $\hat{\cal H} |\Psi_k\rangle = E_k |\Psi_k\rangle$ by assuming $|\Psi_k\rangle = \sum_{i=1}^{5} \sum_{j=1}^{N} c_{i,\kappa} {\rm e}^{{\rm i} \kappa j}|i, j \rangle$. The relative energy of exact one-magnon eigenstates referred with respect to the energy of fully polarized ferromagnetic state $\varepsilon_k = E_k - E_{\rm FM}$ can be calculated from the characteristic equation given by the secular determinant 
\begin{widetext}
\begin{eqnarray}
\left|
\begin{array}{ccccc}
h -4J_1-\varepsilon_k & \frac{J_1}{\sqrt{2}} (1 + {\rm e}^{-{\rm i} \kappa}) & \frac{J_1}{\sqrt{2}} (1 + {\rm e}^{-{\rm i} \kappa}) 
& \frac{J_1}{\sqrt{2}} (1 + {\rm e}^{-{\rm i} \kappa}) & \frac{J_1}{\sqrt{2}} (1 + {\rm e}^{-{\rm i} \kappa}) \\ 	
\frac{J_1}{\sqrt{2}} (1 + {\rm e}^{{\rm i} \kappa}) & h -2J_1-J_2-\varepsilon_k & \frac{J_2}{2} & 0 & \frac{J_2}{2} \\
\frac{J_1}{\sqrt{2}} (1 + {\rm e}^{{\rm i} \kappa}) & \frac{J_2}{2} & h -2J_1-J_2-\varepsilon_k & \frac{J_2}{2} & 0 \\
\frac{J_1}{\sqrt{2}} (1 + {\rm e}^{{\rm i} \kappa}) & 0 & \frac{J_2}{2} & h -2J_1-J_2-\varepsilon_k & \frac{J_2}{2} \\
\frac{J_1}{\sqrt{2}} (1 + {\rm e}^{{\rm i} \kappa}) & \frac{J_2}{2} & 0 & \frac{J_2}{2} & h -2J_1-J_2-\varepsilon_k \\
\end{array}
\right| = 0.
\label{det}
\end{eqnarray}
\end{widetext}
As a result, one gets five one-magnon energy bands of the mixed spin-1 and spin-1/2 Heisenberg octahedral chain 
\begin{eqnarray}
\varepsilon_{1}   \!&=&\! h - 2J_1 - 2J_2,  \nonumber \\
\varepsilon_{2,3} \!&=&\! h -2J_1 -  J_2,  \nonumber \\
\varepsilon_{4,5} \!&=&\! h - J_1 \left(3 \pm \sqrt{5 + 4 \cos \kappa} \right).  
\label{oms}
\end{eqnarray}
A dependence of the one-magnon bands (\ref{oms}) on the wave-vector $\kappa$ is illustrated in Fig. \ref{fig3} at the saturation field for several values of the interaction ratio $J_2/J_1$. It is worth noticing that three one-magnon bands (\ref{oms}) are completely flat (dispersionless), which indicates a bound (localized) character of magnons  within these flat bands.\cite{derz06,derz15} The flat band with the energy $\varepsilon_{1}$ has the lowest energy among all one-magnon eigenstates (\ref{oms}) in the frustrated region $J_2 > 2 J_1$ (see Fig. \ref{fig3}), where a single magnon is preferentially trapped on an elementary square plaquette
\begin{eqnarray}
|lm\rangle_j = \frac{1}{2} \left(\hat{S}_{2,j}^{-} - \hat{S}_{3,j}^{-} + \hat{S}_{4,j}^{-} - \hat{S}_{5,j}^{-}\right) |{\rm FM}\rangle.
\label{om1}
\end{eqnarray}

\begin{figure*}
\begin{center}
\includegraphics[width=0.75\textwidth]{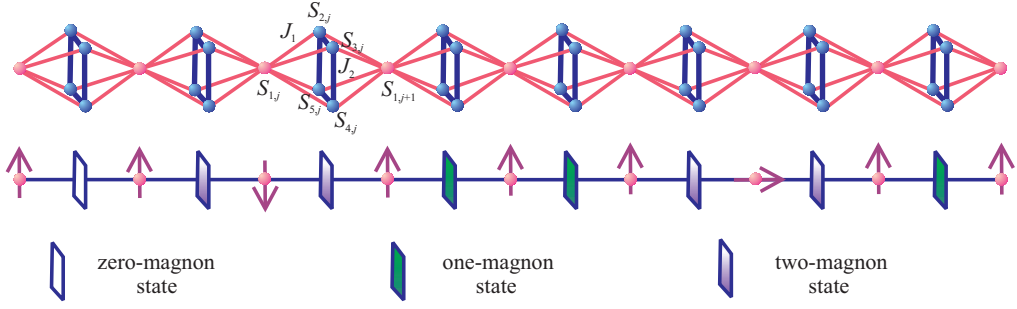}
\vspace{-0.2cm}
\caption{(Color online) A schematic representation of the mixed spin-1 and spin-1/2 Heisenberg octahedral chain and the equivalent two-component lattice-gas model of hard-core monomers valid in a highly frustrated region $J_2/J_1>3$. Black (green) and shaded (violet) parallelograms denote hard-core monomers, which represent one-magnon and two-magnon states of a square plaquette. Unoccupied white parallelogram denotes fully polarized (zero-magnon) state of a square plaquette.}
\label{fig4}
\end{center}
\end{figure*}

It is quite obvious from the previous argumentation that one may construct exact many-magnon eigenstates of the mixed spin-1 and spin-1/2 Heisenberg octahedral chain because the one-magnon eigenstate (\ref{om1}) is locally bound to an elementary square plaquette. Owing to this fact, 
 the localized magnons of the type (\ref{om1}) can be independently placed on square plaquettes, whereas the corresponding many-magnon eigenstates including $r$ bound one-magnon states (\ref{om1}) will have the energy $E_{r} = E_{\rm FM} - r (|\varepsilon_{1}| - h)$. This fact determines in the frustrated region $J_2 > 2 J_1$ a lower bound for the saturation field, because such bound many-magnon eigenstates have lower (the same) energy in comparison with the fully polarized ferromagnetic state below (at) the saturation field $h_s = |\varepsilon_{1}| = 2J_1 + 2 J_2$. In the frustrated region $J_2 > 2 J_1$ and magnetic fields $h<h_s$ the magnon-crystal (MC) phase with the most dense packing ($r=N$) of the localized magnons (\ref{om1}) consequently represents the lowest-energy eigenstate (ground state) 
\begin{eqnarray}
|{\rm MC}\rangle = \prod_{j=1}^N \! |1_{1,j}\rangle \!\otimes\! \frac{1}{2}
(|\!\!\downarrow_{2,j}\uparrow_{3,j}\uparrow_{4,j}\uparrow_{5,j}\rangle 
\!\!\!&-&\!\!\!|\!\!\uparrow_{2,j}\downarrow_{3,j}\uparrow_{4,j}\uparrow_{5,j}\rangle \nonumber \\
+|\!\!\uparrow_{2,j}\uparrow_{3,j}\downarrow_{4,j}\uparrow_{5,j}\rangle
\!\!\!&-&\!\!\!|\!\!\uparrow_{2,j}\uparrow_{3,j}\uparrow_{4,j}\downarrow_{5,j}\rangle). \nonumber \\  
\label{LM}
\end{eqnarray}
The bound magnon-crystal phase (\ref{LM}) should manifest itself in a zero-temperature magnetization curve as the intermediate two-thirds plateau, which should exist within the field range $h \in (J_1 + 2J_2, 2J_1 + 2J_2)$ provided that this ground state appears due to a field-driven phase transition from the monomer-tetramer ground state (\ref{MT}).

\subsection{Low-temperature thermodynamics in terms of an effective lattice-gas model} 
\label{lmt}
On the basis of the previous results it could be concluded that the monomer-tetramer (\ref{MT})
 and the bound magnon-crystal (\ref{LM}) phase are being the respective ground states of the mixed spin-1 and spin-1/2 Heisenberg octahedral chain in a highly frustrated parameter region ($J_2>3J_1$) at low and high magnetic fields, respectively. While the bound two-magnon (singlet-tetramer) eigenstate is present at all square plaquettes within the monomer-tetramer ground state (\ref{MT}), the magnon-crystal ground state (\ref{LM}) involves  at all square plaquettes the bound one-magnon eigenstate. 

With this background, our further attention will be aimed at a proper description of low-temperature thermodynamics of the mixed spin-1 and spin-1/2 Heisenberg octahedral chain, which will be elaborated from bound many-magnon eigenstates assuming on square plaquettes either the localized two-magnon or one-magnon states. The localized nature of bound one- and two-magnon eigenstates allows us to reformulate this problem using the language of a classical lattice-gas model. For this purpose, let us introduce the chemical potentials of two kind of particles $\mu_1 = 2J_1 + 2J_2 - h$ and $\mu_2 = 4J_1 + 3J_2 - 2h$, which determine an energy cost associated with a creation of the bound one-magnon and two-magnon eigenstates at a single square plaquette on the fully polarized ferromagnetic background (see Fig. \ref{fig4}). The localized many-magnon eigenstates of the mixed spin-1 and spin-1/2 Heisenberg octahedral chain can be subsequently described by the classical lattice-gas model 
\begin{eqnarray}
{\cal H} \!= E_{\rm FM}^0 \!-\! h\left(\!2N+\sum_{j=1}^N{S_{1,j}^z}\!\right)\!-\mu_1 \sum_{j=1}^{N} n_{1,j} - \mu_2 \sum_{j=1}^{N} n_{2,j}, \nonumber
\label{elm}
\end{eqnarray}
where $E_{\rm FM}^0 = N (4J_1 + J_2)$ represents the zero-field energy of the fully polarized ferromagnetic state and the occupation number $n_{1,j}=0,1$ ($n_{2,j}=0,1$) determines whether or not the $i$th square plaquette is being occupied by the quasiparticle pertinent to the bound one-magnon (two-magnon) eigenstate, respectively. It is worthwhile to remark that one should also consider a hard-core constraint (1-$n_{1,j}n_{2,j}$) for two kinds of quasiparticles, which excludes a double occupancy of square plaquettes by the bound one- and two-magnon eigenstates when calculating the partition function according to the formula
\begin{eqnarray}
{\cal Z} \!\!&=&\!\! {\rm e}^{-\beta E_{\rm FM}^0+2\beta Nh} \prod_{j=1}^N \sum_{S_{1,j}^z} \sum_{n_{1,j}} \sum_{n_{2,j}} (1-n_{1,j}n_{2,j}) \times \nonumber \\
\!\!&\times&\!\!
{\rm e}^{\beta (\mu_1 n_{1,j} + \mu_2 n_{2,j})+\beta h S_{1,j}^z} \nonumber \\
\!\!&=&\!\! {\rm e}^{-\beta E_{\rm FM}^0+2\beta Nh} \!\left(1 \!\!+\! 2\cosh \beta h\right)^N \!\!\left(1 \!+\!{\rm e}^{\beta \mu_1} \!+\! {\rm e}^{\beta \mu_2}\right)^N. \nonumber 
\end{eqnarray}
Here, $\beta = 1/(k_{\rm B} T)$, $k_{\rm B}$ is Boltzmann's constant and $T$ is the absolute temperature. The Helmholtz free energy per elementary unit can be calculated from the relation 
\begin{eqnarray}
f  \!\!&=&\!\! -k_{\rm B} T \lim_{N \to \infty} \frac{1}{N} \ln {\cal Z} \nonumber \\
\!\!&=&\!\! (4J_1 \!+\! J_2) \!-\! 2h\!- k_{\rm{B}}T\ln\left(1+2\cosh\beta h\right) \nonumber \\
\!\!&-&\!\! \!k_{\rm{B}}T\ln\left(1 + {\rm e}^{\beta \mu_1} \!+\! {\rm e}^{\beta \mu_2} \right).
\label{lmgfe}
\end{eqnarray}
The other thermodynamic quantities follow straightforwardly from Eq. (\ref{lmgfe}). For instance, the isothermal magnetization per unit cell is given by
\begin{eqnarray}
m  = \!-\!\left(\frac{\partial f}{\partial h}\right)_{\! T}\!\!\!=\! 2 \!+\! \frac{\sinh \beta h}{1+2\cosh \beta h} \!-\! \frac{\rm{e}^{\beta \mu_1} + 2{\rm{e}^{\beta \mu_2}}}{1+\rm{e}^{\beta \mu_1}+ {\rm{e}^{\beta \mu_2}}}.
\label{magperunit}
\end{eqnarray}

{\
\subsection{DMRG simulations of the effective mixed-spin Heisenberg chains} 
\label{dmrg}

The Hamiltonian (\ref{ham}) of the mixed spin-1 and spin-1/2 Heisenberg octahedral chain can be re-expressed in terms of the composite spin operator of a square plaquette $\boldsymbol{\hat{S}}_{\square, j} = \boldsymbol{\hat{S}}_{2,j} + \boldsymbol{\hat{S}}_{3,j} + \boldsymbol{\hat{S}}_{4,j} + \boldsymbol{\hat{S}}_{5,j}$ and the composite spin operators $\boldsymbol{\hat{S}}_{24, j} = \boldsymbol{\hat{S}}_{2,j} + \boldsymbol{\hat{S}}_{4,j}$, $\boldsymbol{\hat{S}}_{35, j} = \boldsymbol{\hat{S}}_{3, j} + \boldsymbol{\hat{S}}_{5, j}$ of two spin pairs from opposite corners of a square plaquette (see Figs. \ref{fig1} and \ref{fig2})
\begin{eqnarray}
\hat{\cal H} \!\!&=&\!\! J_1 \sum_{j=1}^{N} (\boldsymbol{\hat{S}}_{1,j} + \boldsymbol{\hat{S}}_{1,j+1}) \!\cdot\! \boldsymbol{\hat{S}}_{\square, j} - h \sum_{j=1}^{N} 
(\hat{S}_{1,j}^{z} + \hat{S}_{\square, j}^z)\nonumber \\  
\!\!&+&\!\! \frac{J_2}{2} \sum_{j=1}^{N} (\boldsymbol{\hat{S}}_{\square, j}^2 - \boldsymbol{\hat{S}}_{24, j}^2  - \boldsymbol{\hat{S}}_{35, j}^2).  
\label{hamlcl}  
\end{eqnarray}
It can be readily proved that the Hamiltonian (\ref{hamlcl}) commutes with a square of all three introduced composite spin operators $[\hat{\cal H}, \boldsymbol{\hat{S}}_{\square, j}^2] = [\hat{\cal H},\boldsymbol{\hat{S}}_{24, j}^2] = [\hat{\cal H}, \boldsymbol{\hat{S}}_{35, j}^2] = 0$, which consequently represent locally conserved quantities with well defined quantum spin numbers (see Appendix~A for a rigorous mathematical proof). Although the Hamiltonian (\ref{hamlcl}) is exactly equivalent to the original Hamiltonian (\ref{ham}) of the mixed spin-1 and spin-1/2 Heisenberg octahedral chain, it also allows derivation of several approximate effective Heisenberg spin models when considering a particular combination of the quantum spin numbers $S_{\square,j}$, $S_{24,j}$ and $S_{35,j}$ substantially reducing a computational complexity (dimension of the Hilbert space) of the studied problem. Indeed, the first two terms of the Hamiltonian (\ref{hamlcl}) correspond to a ferrimagnetic mixed spin-$(1,S_{\square,j})$ Heisenberg chain, while the last term provides just a trivial shift of the energy depending on a size of chosen quantum spin numbers $S_{\square,j}$, $S_{24,j}$ and $S_{35,j}$. The ground state of the mixed spin-1 and spin-1/2 Heisenberg octahedral chain can be accordingly found from the lowest-energy eigenstates of the effective mixed spin-($S_{1,j}$, $S_{\square,j}$) Heisenberg chains unambiguously given by the Hamiltonian (\ref{hamlcl}) under the specific choice of the quantum spin numbers $S_{\square,j}$, $S_{24,j}$ and $S_{35,j}$. Note that the total spin on a square plaquette may achieve three different values $S_{\square,j} = 0$, $1$, and $2$, whereas the former value $S_{\square,j} = 0$ corresponding to a singlet-tetramer state is being responsible for a fragmentation of the effective mixed-spin Heisenberg chains. The lowest-energy eigenstates of the effective mixed spin-($S_{1,j}$, $S_{\square,j}$) Heisenberg chains, which involve the singlet-plaquette state $S_{\square,j} = 0$ at periodic positions, can be thus readily calculated by exact analytical or numerical diagonalization of smaller Heisenberg spin clusters. On the other hand, one has to resort to some powerful numerical technique such as DMRG method in order to obtain the lowest-energy eigenstates of the effective mixed spin-($S_{1,j}$, $S_{\square,j}$) Heisenberg chains when considering the states with $S_{\square,j} = 1$ and/or $2$. Adapting subroutines from ALPS project \cite{baue11} we have therefore performed DMRG simulations for several effective spin-($S_{1,j}$, $S_{\square,j}$) Heisenberg chains with the translational period less than four and the total number of spins $120$, which are equivalent to the large-scale DMRG simulations of the mixed spin-1 and spin-1/2 Heisenberg octahedral chain with the total number of spins 300.

\subsection{Lanczos method and full ED calculations}

A substantial reduction of the dimension of the Hilbert space achieved in Sect. \ref{dmrg} is however balanced by incapability of performing DMRG simulations for the effective Hamiltonians (\ref{hamlcl}) of the mixed spin-($S_{1,j}$, $S_{\square,j}$) Heisenberg chains with all possible combinations of the quantum spin numbers $S_{\square,j}$, $S_{24,j}$ and $S_{35,j}$. To avoid danger of overlooking of some ground states we have alternatively performed an exact diagonalization for the original Hamiltonian (\ref{ham}) of a finite-size mixed spin-1 and spin-1/2 Heisenberg octahedral chain being composed of four unit cells $N=4$ (20 spins) by implementing the Lanczos algorithm from ALPS project.\cite{baue11} 

Similarly, the effective lattice-gas model derived in Sect. \ref{lmt} for a description of the low-temperature thermodynamics of the mixed spin-1 and spin-1/2 Heisenberg octahedral chain in a highly frustrated parameter region is just approximate and hence, one also has to prove its reliability. For this purpose, thermodynamic analysis of the effective two-component lattice-gas model  was always confronted with the exact numerical diagonalization of finite-size mixed spin-1 and spin-1/2 Heisenberg octahedral chain being composed of three unit cells $N=3$ (15 spins). This comparison has allowed us to determine a temperature range over which the effective lattice-gas model provides a plausible description of thermodynamic properties of the original model. 
}

\section{Results and discussion}
\label{sec:result}

This section will be devoted to a detailed analysis of the most interesting results, which have been obtained for the ground state, magnetization curves and  low-temperature thermodynamics of the mixed spin-1 and spin-1/2 Heisenberg octahedral chain within the help of methods thoroughly described in a previous section.    

{\
\subsection{Ground-state phase diagrams and zero-temperature magnetization curves}

The overall ground-state phase diagram of the mixed spin-1 and spin-1/2 Heisenberg octahedral chain is displayed in Fig. \ref{fig5} in the $J_2/J_1 - h/J_1$ plane as obtained from the DMRG simulations of the effective mixed spin-($S_{1,j}$, $S_{\square,j}$) Heisenberg chains supplemented with exact calculations. It turns out that the investigated quantum spin chain already displays a plethora of exotic quantum phases at zero magnetic field. At relatively small values of the interaction ratio $J_2/J_1<1.018$ the ground state of the mixed spin-1 and spin-1/2 Heisenberg octahedral chain is the Lieb-Mattis ferrimagnetic phase, which originates from the effective mixed spin-($1,2$) Heisenberg chain \cite{ivanov98} with the highest possible value of the composite spin $S_{\square,j} = 2$ on all square plaquettes. The other ferrimagnetic phase with a doubled period of the magnetic unit cell ($p=2$) can be only found in a relatively narrow interval of the parameter space $J_2/J_1 \in (1.018, 1.073)$. This ground state follows from the lowest-energy eigenstate of the effective mixed spin-($1,1,1,2$) Heisenberg chain with a regular alternation of the triplet ($S_{\square,j}=1$) and quintet ($S_{\square,j}=2$) states on odd and even square plaquettes (or vice versa). It is quite plausible to conjecture that this ground state also belongs to a class of Lieb-Mattis ferrimagnetic states although the four-sublattice character of the effective mixed spin-($1,1,1,2$) Heisenberg chain precludes the simple argumentation on the grounds of Lieb-Mattis theorem.\cite{lieb62} 

\begin{figure}
\begin{center}
\hspace{-1.3cm}
\includegraphics[width=0.555\textwidth]{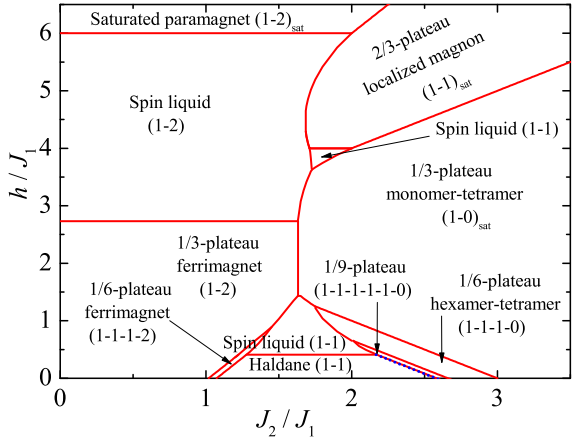}
\end{center}
\vspace{-1.5cm}
\caption{(Color online) The ground-state phase diagram of the mixed spin-1 and spin-1/2 Heisenberg octahedral chain in the $J_2/J_1 - h/J_1$ plane. The dotted line delimits a phase boundary along which a tiny one-twelfth plateau related to the cluster-based Haldane phase with $p=4$ appears in a narrow range of the magnetic fields. The numbers in brackets determine the total spin of monomeric sites and square plaquettes within a magnetic unit cell of a given ground state.}
\label{fig5}
\end{figure}

The uniform Haldane phase represents an exact ground state in the parameter region $J_2/J_1 \in (1.073,2.577)$ and this ground state can be descended from the lowest-energy eigenstate of the effective spin-(1,1) Heisenberg chain with the composite spin $S_{\square,j} = 1$ on all square plaquettes. Most strikingly, one also encounters in a relatively narrow range of the interaction ratio $J_2/J_1 \in (2.577,2.583)$ and $J_2/J_1 \in (2.583,2.660)$ two related cluster-based Haldane phases, which have higher period of the magnetic unit cell $p=4$ and $p=3$, respectively. In contrast to the uniform Haldane phase, the fragmentized cluster-based Haldane phases disturb the translational symmetry because of a periodic repetition of the plaquette-singlet state $S_{\square,j}=0$ breaking the octahedral chain into smaller fragments. It is noteworthy that the stability of the fragmentized cluster-based Haldane phases is inversely proportional to a period of the magnetic unit cell $p$. In fact, the hexamer-tetramer ground state with a regular alternation of the triplet-hexamer and singlet-tetramer states, as the last member of this family with the specific period $p=2$, extends over a much wider interval of the parameter space $J_2/J_1 \in (2.660,3)$ than other two fragmentized cluster-based Haldane phases together. The hexamer-tetramer ground state originates from the effective mixed spin-($1,1,1,0$) Heisenberg chain with a regular alternation of composite triplet ($S_{\square,j}=1$) and singlet ($S_{\square,j}=0$) states on odd and even square plaquettes (or vice versa). Last but not least, the monomer-tetramer ground state (\ref{MT}) with the composite singlet state $S_{\square,j}=0$ on all square plaquettes emerges in a highly-frustrated parameter region $J_2/J_1>3$ in agreement with the variational arguments presented in Sect. \ref{vm}. The monomeric spin-1 particles become within the monomer-tetramer ground state (\ref{MT}) paramagnetic due to absence of spin-spin correlations across 
the singlet-tetramer state. 

At this stage, let us provide a more comprehensive understanding of unconventional cluster-based Haldane phases, which are for better illustration schematically depicted in Fig. \ref{fig6} along with the monomer-tetramer and uniform Haldane phases. The shaded ovals represent the lowest-energy triplet state of the finite-size spin-1 Heisenberg chain under open boundary condition as the underlying spin cluster, which is isolated at both its sides through the singlet-plaquette state. It should be pointed out that just three cluster-based Haldane phases with the magnetic period $p=2$, $3$ and $4$ may represent true ground states of the mixed spin-1 and spin-1/2 Heisenberg octahedral chain (see Appendix B). This result is in sharp contrast to an infinite series of ground states caused by fragmentation reported by Schulenburg and Richter for the spin-1/2 Heisenberg orthogonal-dimer chain.\cite{schu02,schul02,johannes} To be more specific, the hexamer-tetramer ground state as the simplest representative of the cluster-based Haldane phase with the magnetic period $p=2$ corresponds to a regular alternation of the plaquette singlet and octahedral (hexamer) triplets [see Fig. \ref{fig6}(b)]. Moreover, the triplet state of the mixed spin-1 and spin-1/2 Heisenberg octahedron (Fig. \ref{fig2}) emergent within the hexamer-tetramer ground state is quite analogous to the cluster-based Haldane state recently reported by Fujihala \textit{et al}. for the mineral-crystal fedotovite.\cite{fuji18} 

\begin{figure*}
\begin{center}
\includegraphics[width=0.6\textwidth]{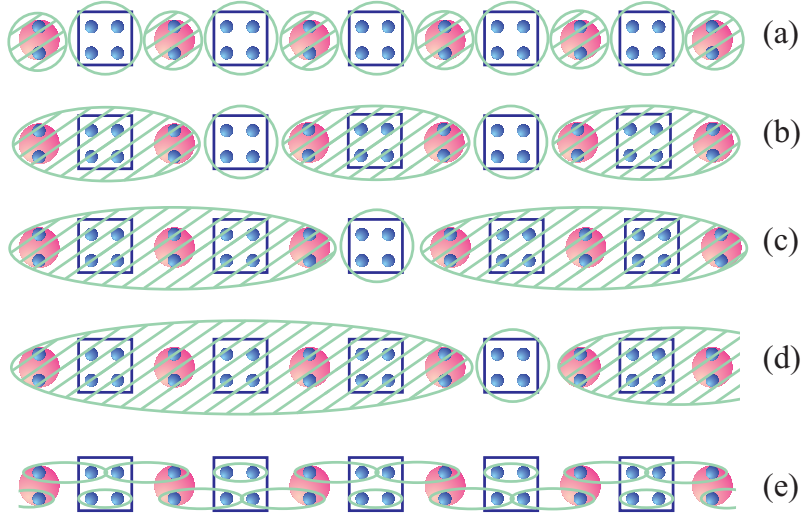}
\end{center}
\vspace{-0.6cm}
\caption{(Color online) A schematic representation of: (a) the  monomer-tetramer phase; (b) hexamer-tetramer state as the simplest cluster-based Haldane state with the period $p=2$;  
(c)-(d) the fragmentized cluster-based Haldane states with the period $p=3$ and 4; (e) the uniform Haldane state. Solid ovals and circles represent singlet-dimer and singlet-tetramer states, respectively, while shaded circles and ovals denote triplet states of a given cluster.}
\label{fig6}
\end{figure*}

\begin{figure*}
\begin{center}
\includegraphics[width=0.99\textwidth]{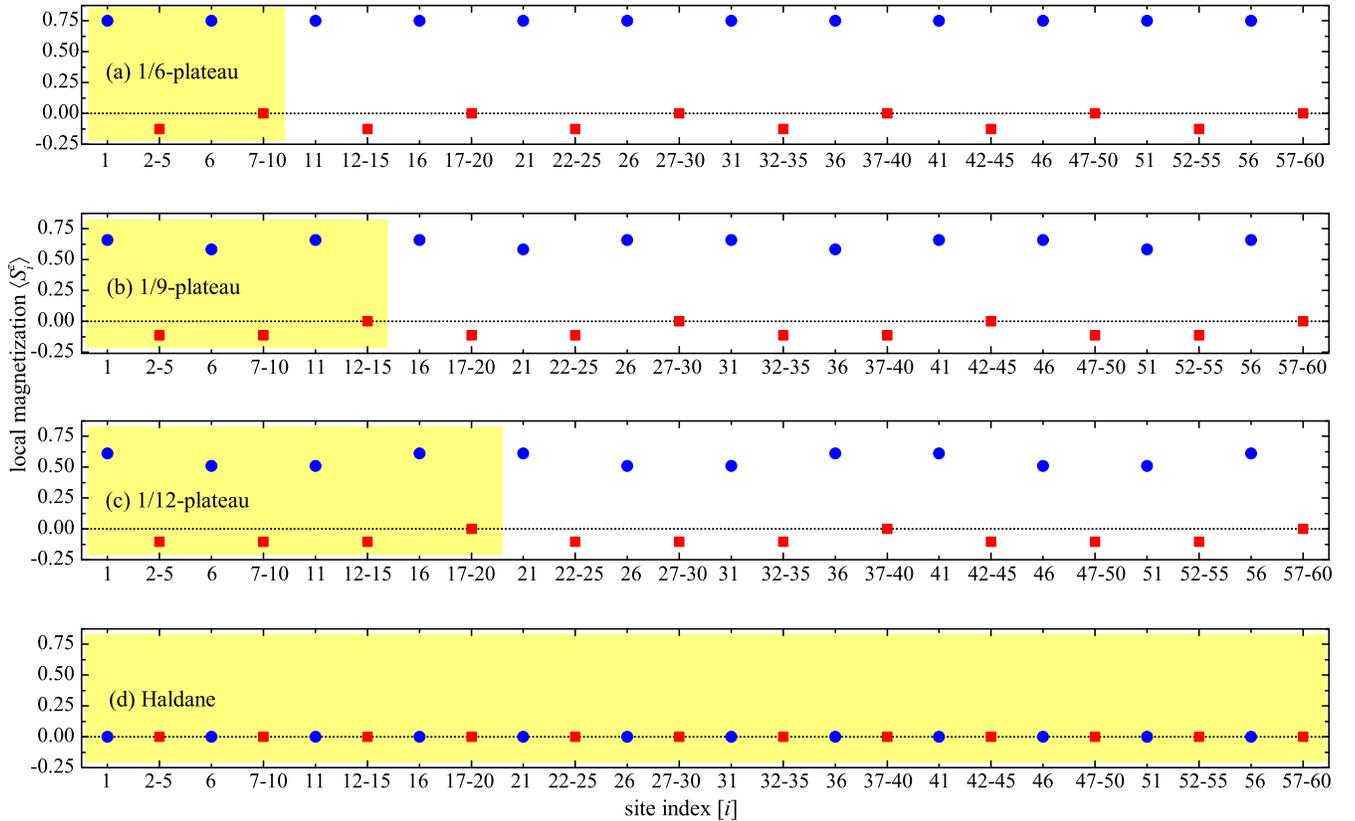}
\end{center}
\vspace{-1.4cm}
\caption{(Color online) Local magnetizations as obtained from DMRG simulations of the mixed spin-1 and spin-1/2 Heisenberg octahedral chain with $N=12$ unit cells (i.e. 60 spins) within: (a) one-sixth plateau (hexamer-tetramer phase with $p=2$), (b) one-ninth plateau ($p=3$), (c) one-twelfth plateau ($p=4$), (d) zero plateau (the uniform Haldane state with $p=\infty$). Each square represents a local magnetization of four spin-1/2 particles from a square  plaquette, while each circle represents local magnetizations of the monomeric spin-1 particle. The shaded space denotes magnetic unit cell.}
\label{fig7}
\end{figure*}

To provide an independent check of the character of the cluster-based Haldane phases we have depicted in Fig. \ref{fig7} local magnetizations as a function of site numbering within this peculiar class of fragmentized ground states, which should manifest itself in zero-temperature magnetization curves as intermediate plateaus at 1/3p of the saturation magnetization. It should be noticed that the four spins belonging to the same square plaquette have the same local magnetization and hence, these local magnetizations were merged together. All local magnetizations of the spin-1/2 particles within the singlet-plaquette state are of course equal zero due to its non-magnetic nature. In the hexamer-tetramer phase all monomeric spin-1 particles have the same local magnetization $\langle \hat{S}_{1,i}^z\rangle=0.75$, while local magnetizations of four spins from the square plaquettes alternate between finite negative value $\langle \hat{S}_{j, 2i-1}^z \rangle = -0.125$ and zero  value $\langle \hat{S}_{j, 2i}^z \rangle =0$ ($j=2,3,4,5$) as exemplified in Fig. \ref{fig7}(a). As one can see from Fig. \ref{fig7}(b) and (c), the local magnetization of the monomeric spins-1 particles belonging to a finite cluster in a triplet state is different inside of this cluster and at an interface with the plaquette-singlet state. The local magnetization of the monomeric spin-1 particles at edges with the plaquette-singlet states is higher than the local magnetization of the monomeric spins inside of the cluster. This difference could signal tendency for a formation of edge states, which are however spread over a few lattice sites (the correlation length) within the Haldane-type phase \cite{kenz03} in contrast to strictly localized edge states of the AKLT model.\cite{affl87,affl88} 
}

Next, our attention will be focused on a ground-state analysis at finite magnetic fields. First, we will review the magnetization values of all aforementioned zero-field ground states.
The two quantum ferrimagnetic ground states related to the lowest-energy eigenstates of the effective mixed spin-($1,2$) and spin-($1,1,1,2$) Heisenberg chains should manifest themselves in zero-temperature magnetization curves as intermediate plateaus at one-third and one-sixth of the saturation magnetization, respectively. Contrarily, the uniform Haldane phase should be responsible for a zero magnetization plateau, while three fragmentized cluster-based Haldane phases with the period $p=2, 3$ and $4$ should cause one-sixth, one-ninth and one-twelfth plateau, respectively. The monomer-tetramer ground state (\ref{MT}) affords another one-third plateau, which solely arises from a full polarization of the monomeric 
spin-1 particles. 

\begin{figure*}
\begin{center}
\includegraphics[width=0.45\textwidth]{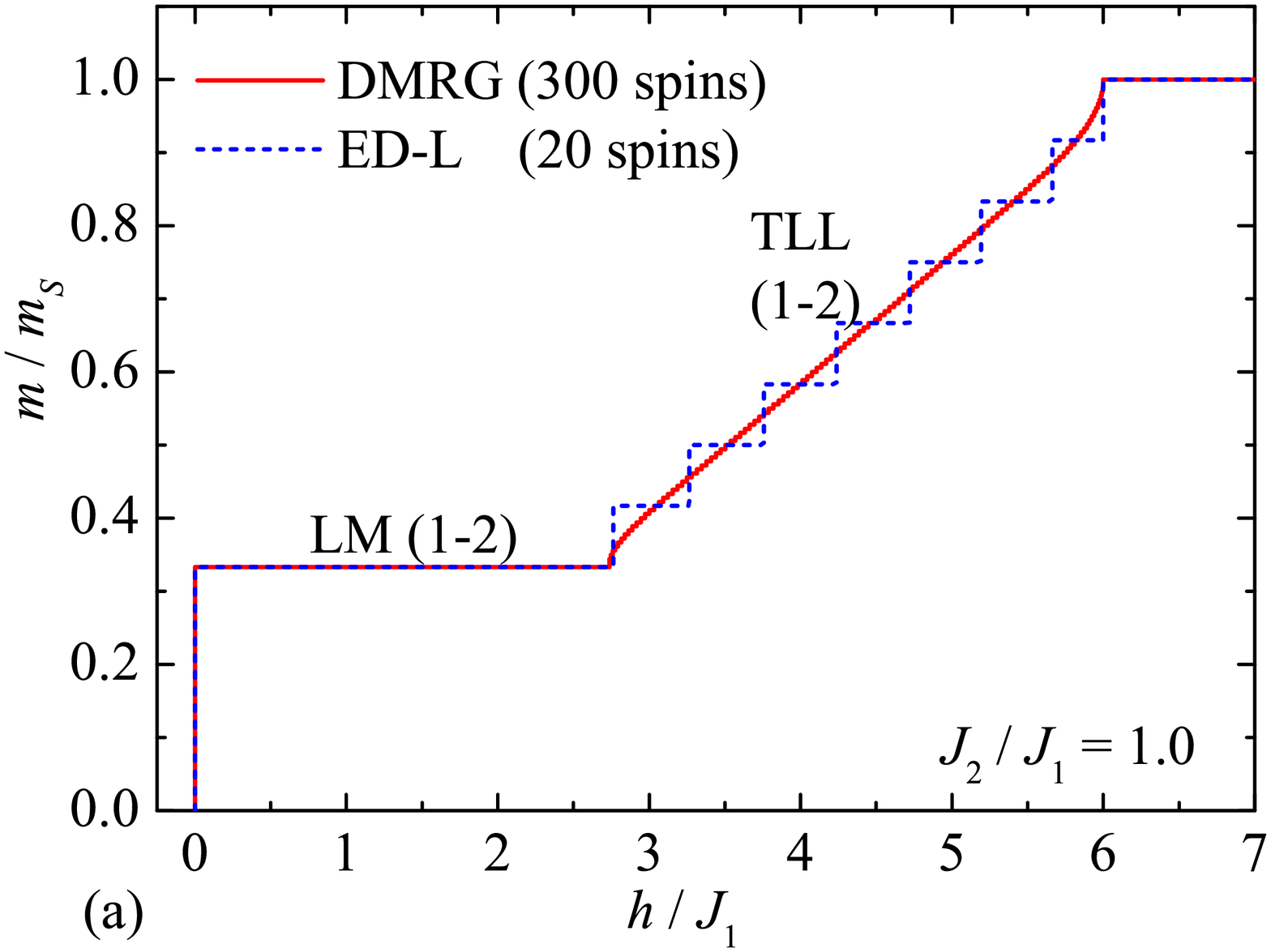}
\hspace{0.5cm}
\includegraphics[width=0.45\textwidth]{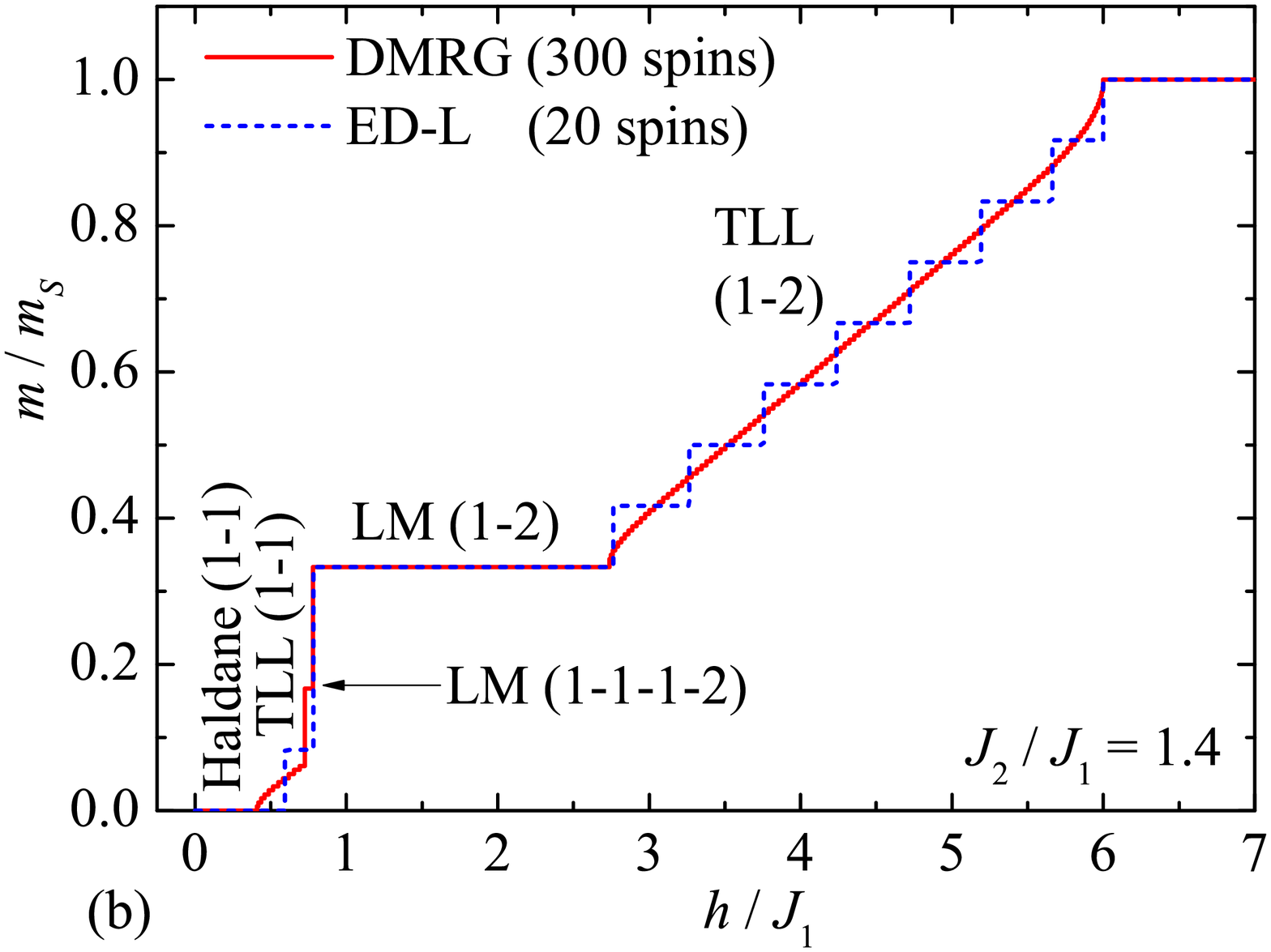}
\includegraphics[width=0.45\textwidth]{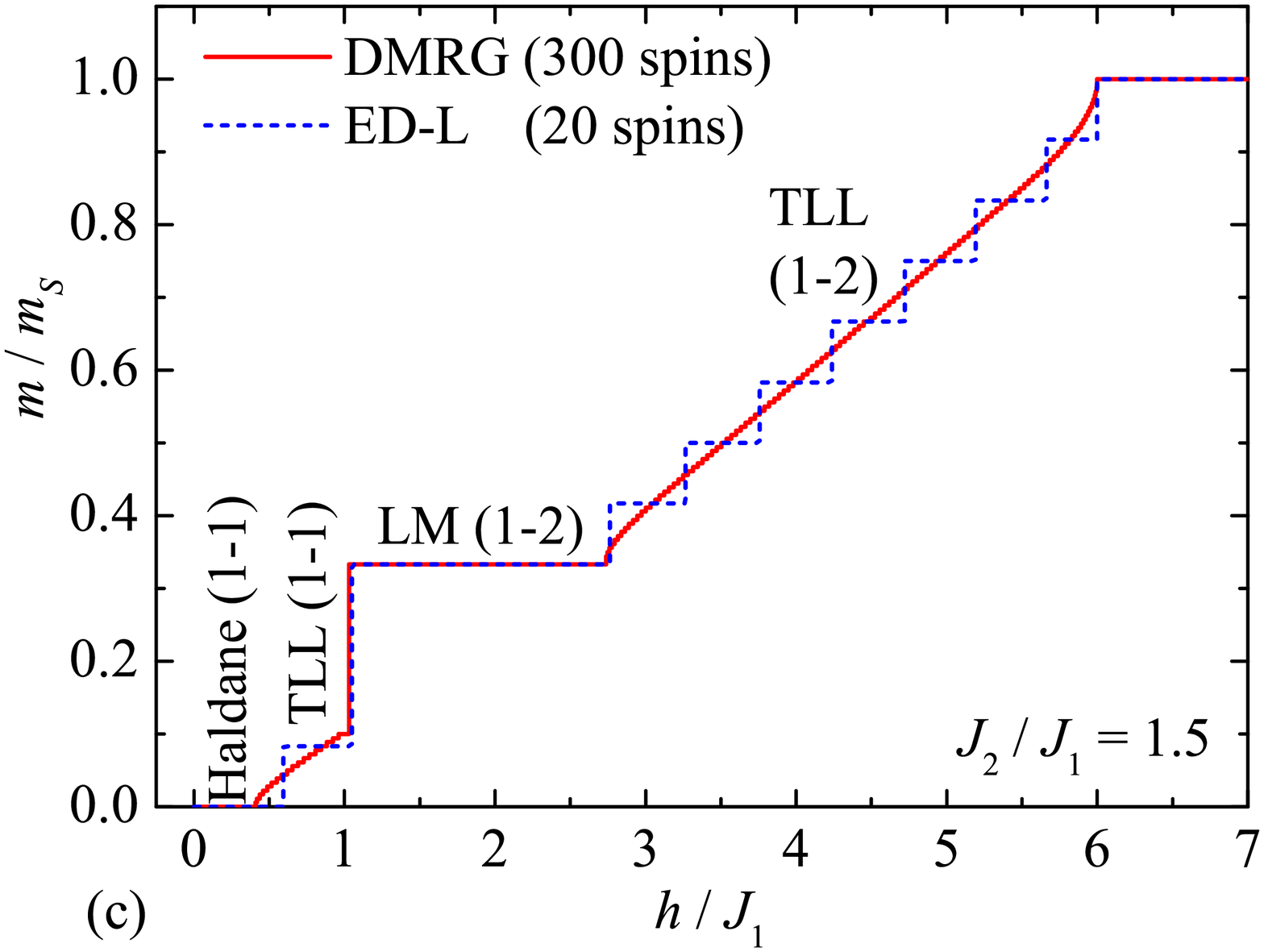}
\hspace{0.5cm}
\includegraphics[width=0.45\textwidth]{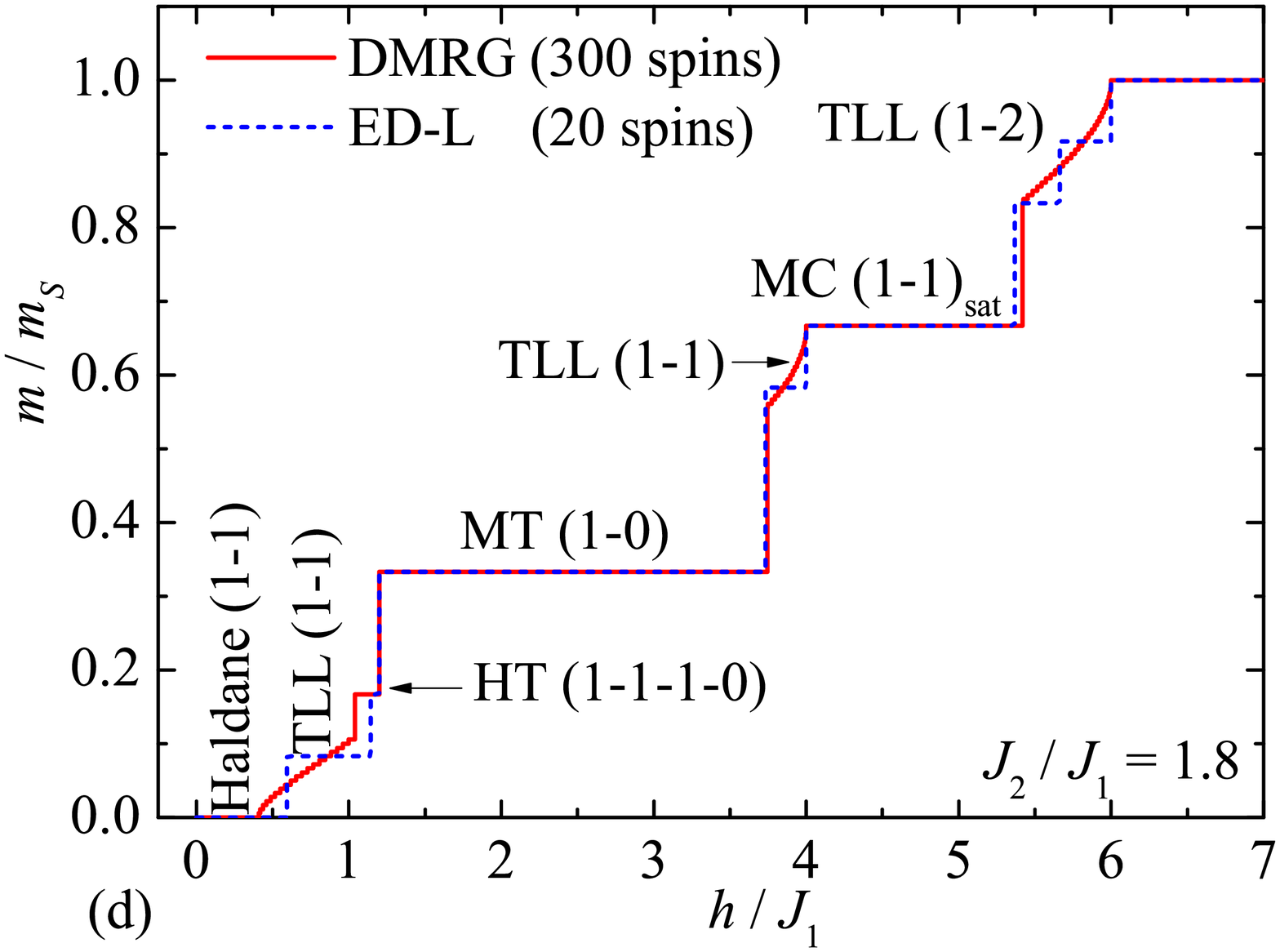}
\includegraphics[width=0.45\textwidth]{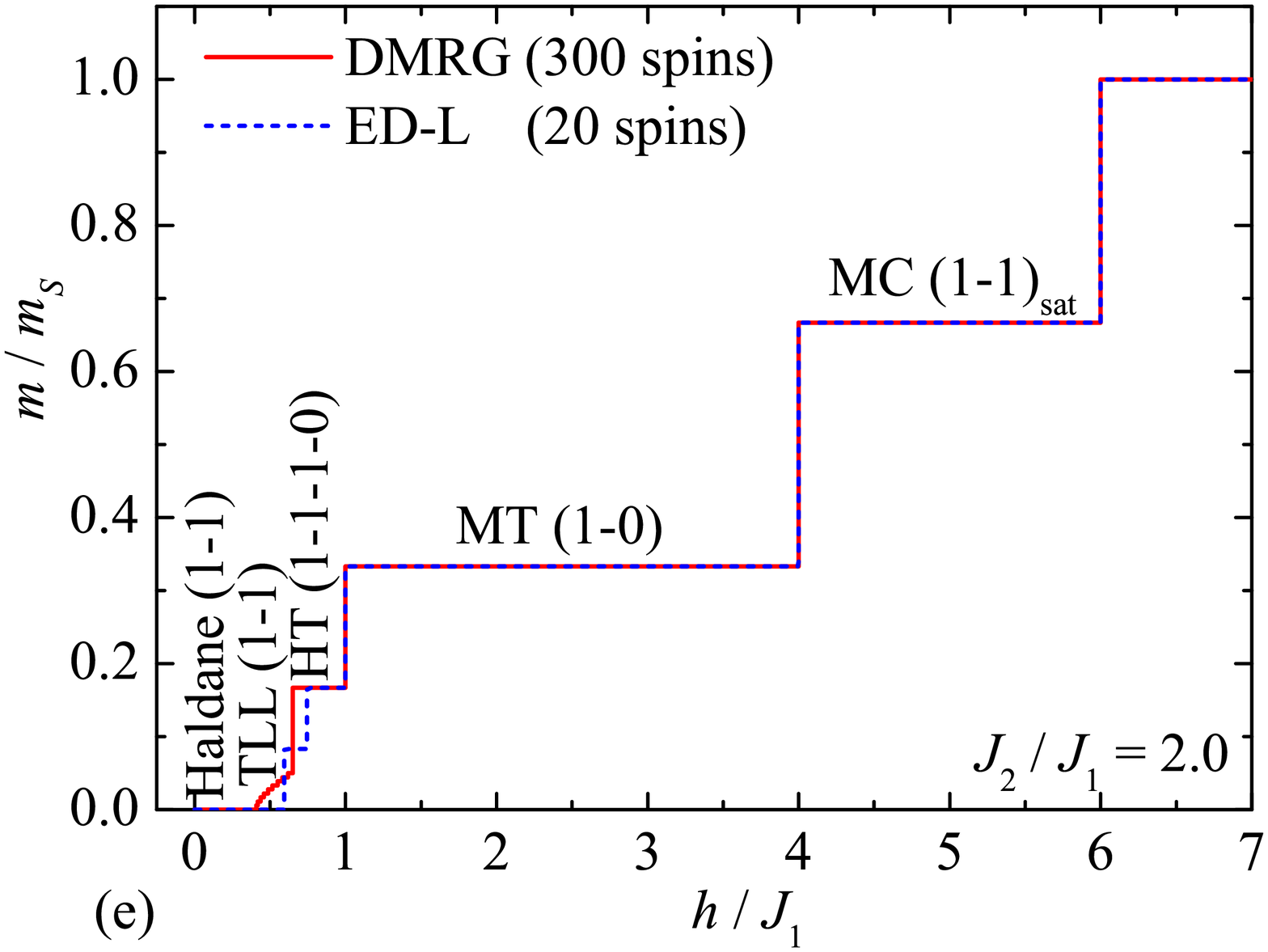}
\hspace{0.5cm}
\includegraphics[width=0.45\textwidth]{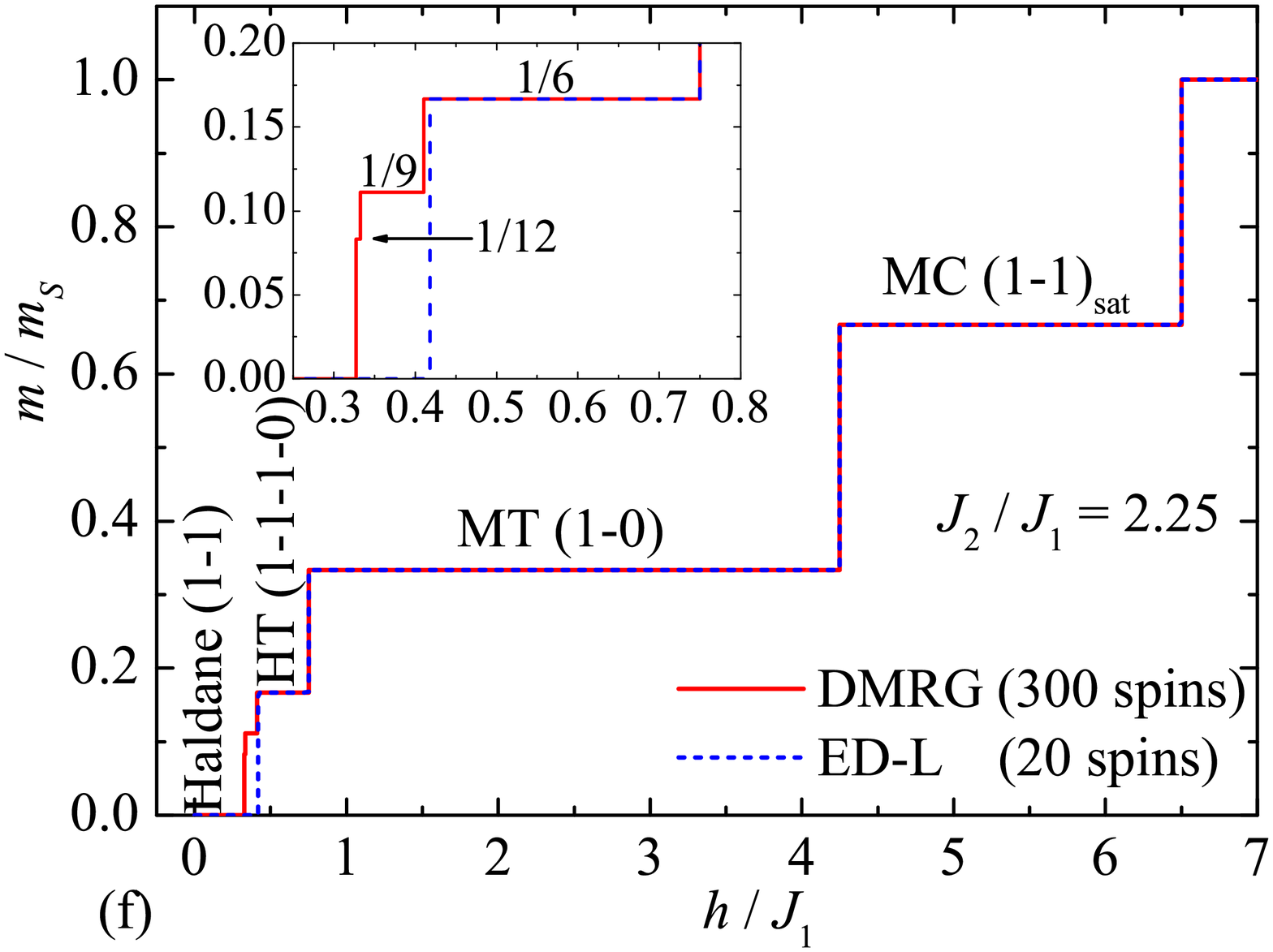}
\end{center}
\vspace{-1.0cm}
\caption{(Color online) Zero-temperature magnetization curves of the mixed spin-1 and spin-1/2 Heisenberg octahedral chain. Solid lines display DMRG  simulations of the effective  mixed-spin Heisenberg chains composed of $N=60$ unit cells (corresponding to $300$ spins), while broken lines display ED data obtained from Lanczos technique for finite-size mixed-spin Heisenberg octahedral chain with $N=4$ unit cells ($20$ spins). To illustrate overall diversity the magnetization curves are plotted for six selected values of the interaction ratio: (a) $J_2/J_1 = 1.0$; (b) $J_2/J_1 = 1.2$; (c) $J_2/J_1 = 1.4$; (d) $J_2/J_1 = 1.5$; (e) $J_2/J_1 = 2.0$; (f) $J_2/J_1 = 2.25$. The inset in Fig. \ref{fig8}(f) shows in an enlarged scale the parameter region, where two tiny  one-ninth and one-twelfth plateaus due to the cluster-based Haldane phases with the period $p=3$ and 4 appear.} 
\label{fig8}
\end{figure*}

The ground-state phase diagram depicted in Fig. \ref{fig5} is especially diverse in a less frustrated parameter space $J_2/J_1 < 3$ with regard to existence of three quantum spin-liquid regions, two of which come from the effective spin-(1,1) Heisenberg chain and third one resulting from the effective mixed spin-(1,2) Heisenberg chain. The ground state of the mixed spin-1 and spin-1/2 Heisenberg octahedral chain originates exclusively from the lowest-energy eigenstates of the effective mixed spin-($1,2$) Heisenberg chain when the coupling ratio is weaker than $J_2/J_1<1.018$. The intermediate one-third plateau related to the Lieb-Mattis ferrimagnetic phase accordingly terminates at a field-induced quantum phase transition towards the gapless Tomonaga-Luttinger spin-liquid ground state emergent at the critical field $h_c/J=2.733$ [Fig. \ref{fig8}(a)]. If the ratio of the coupling constants is from the interval $J_2/J_1 \in (1.018, 1.073)$ one obtains the analogous  zero-temperature magnetization curve with only one exception that a tiny one-sixth plateau pertinent to the ferrimagnetic phase with the doubled period of the magnetic unit cell is present prior to the more extensive one-third plateau. 

The zero-magnetization plateau related to the uniform Haldane phase emerges when the interaction ratio is selected from the interval $J_2/J_1 \in (1.073,2.577)$, see Figs. \ref{fig8}(b)-(f). The uniform Haldane phase either breaks down at a discontinuous field-driven quantum phase transition towards the ferrimagnetic phase with a doubled period (one-sixth plateau) or a continuous field-driven quantum phase transition towards the Tomonaga-Luttinger quantum spin liquid [see Fig. \ref{fig8}(b)-(e)] or a discontinuous field-driven phase transition towards the fragmentized cluster-based Haldane phase with the period $p=4$ corresponding to one-twelfth plateau [see Fig. \ref{fig8}(f)]. 

It should be pointed out that the microscopic nature of a wide one-third plateau basically depends on whether the interaction ratio is smaller or greater than $J_2/J_1=1.631$. In the former case $J_2/J_1<1.631$  the upper critical field associated with the breakdown of one-third plateau of Lieb-Mattis type is independent of the coupling ratio $J_2/J_1$, while the upper critical field of one-third plateau relevant to the monomer-tetramer phase (\ref{MT}) monotonically increases towards higher magnetic field upon strengthening of the interaction ratio  $J_2/J_1$ [c.f. Figs. \ref{fig8}(a)-(c) with Fig. \ref{fig8}(d)]. Owing to this fact, the magnetic field region inherent to the quantum spin liquid of the effective mixed spin-(1,2) Heisenberg chain substantially shrinks whenever $J_2/J_1>1.631$. The other quantum spin liquid arising from the effective spin-(1,1) Heisenberg chain exhibits a peculiar reentrant behavior when it also appears at higher magnetic fields besides the low-field region closing an energy  gap above the uniform Haldane phase [see Fig. \ref{fig8}(d)].

Zero-temperature magnetization curve involves two-thirds plateau inherent to the bound magnon-crystal phase (\ref{LM}) whenever the interaction ratio exceeds $J_2/J_1=1.685$. The bound magnon-crystal phase is either wedged in between two quantum spin-liquid regions [see Fig. \ref{fig8}(d)] or it appears due to a discontinuous  magnetization jump from the one-third plateau  relevant to the monomer-tetramer phase [see Figs. \ref{fig8}(e) and (f)]. It is noteworthy that the intermediate two-thirds plateau due to the  magnon-crystal state (\ref{LM}) with a single magnon bound on each square plaquette can alternatively be interpreted as the saturated state of the effective spin-(1,1) Heisenberg chain. 

Last but not least, it follows from Fig. \ref{fig5} that the ground-state phase diagram is fully consistent with presence of the monomer-tetramer state (\ref{MT}) and the bound magnon-crystal state (\ref{LM}) predicted in Sects. \ref{vm} and \ref{lmgs} for the highly-frustrated parameter region $J_2/J_1 > 3$  by making use of the variational procedure and the localized-magnon approach, respectively. The localized nature of bound two- and one-magnon states within the monomer-tetramer phase (\ref{MT}) and the magnon-crystal phase (\ref{LM}) additionally allows a classical description of low-temperature magnetothermodynamics using the localized-magnon approach elaborated in Sect. \ref{lmt}, which will be comprehensively examined in the following section. 

\subsection{Magnetothermodynamics in a highly frustrated parameter region}

\begin{figure*}
\begin{center}
\includegraphics[width=0.45\textwidth]{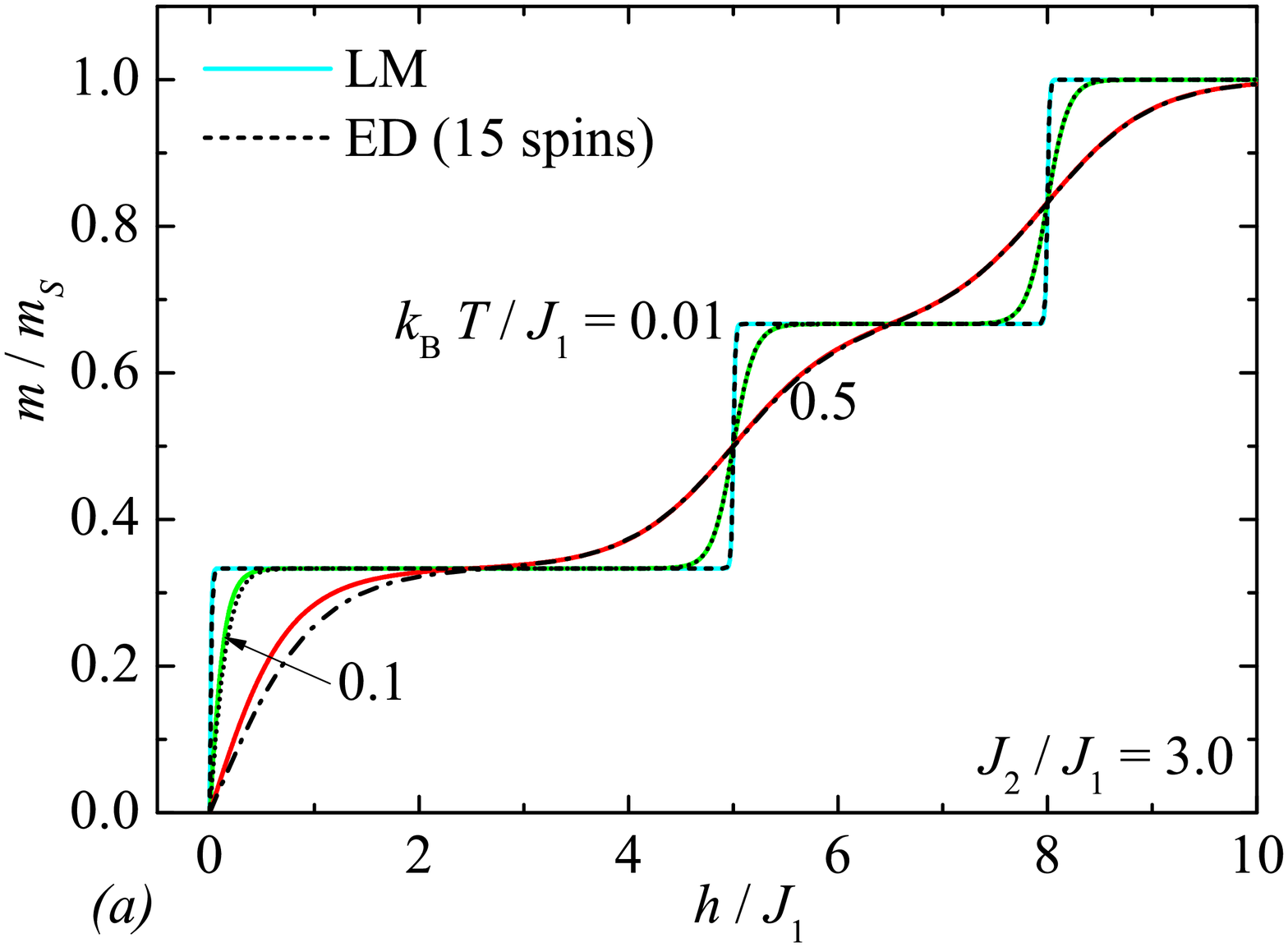}
\hspace{0.5cm}
\includegraphics[width=0.45\textwidth]{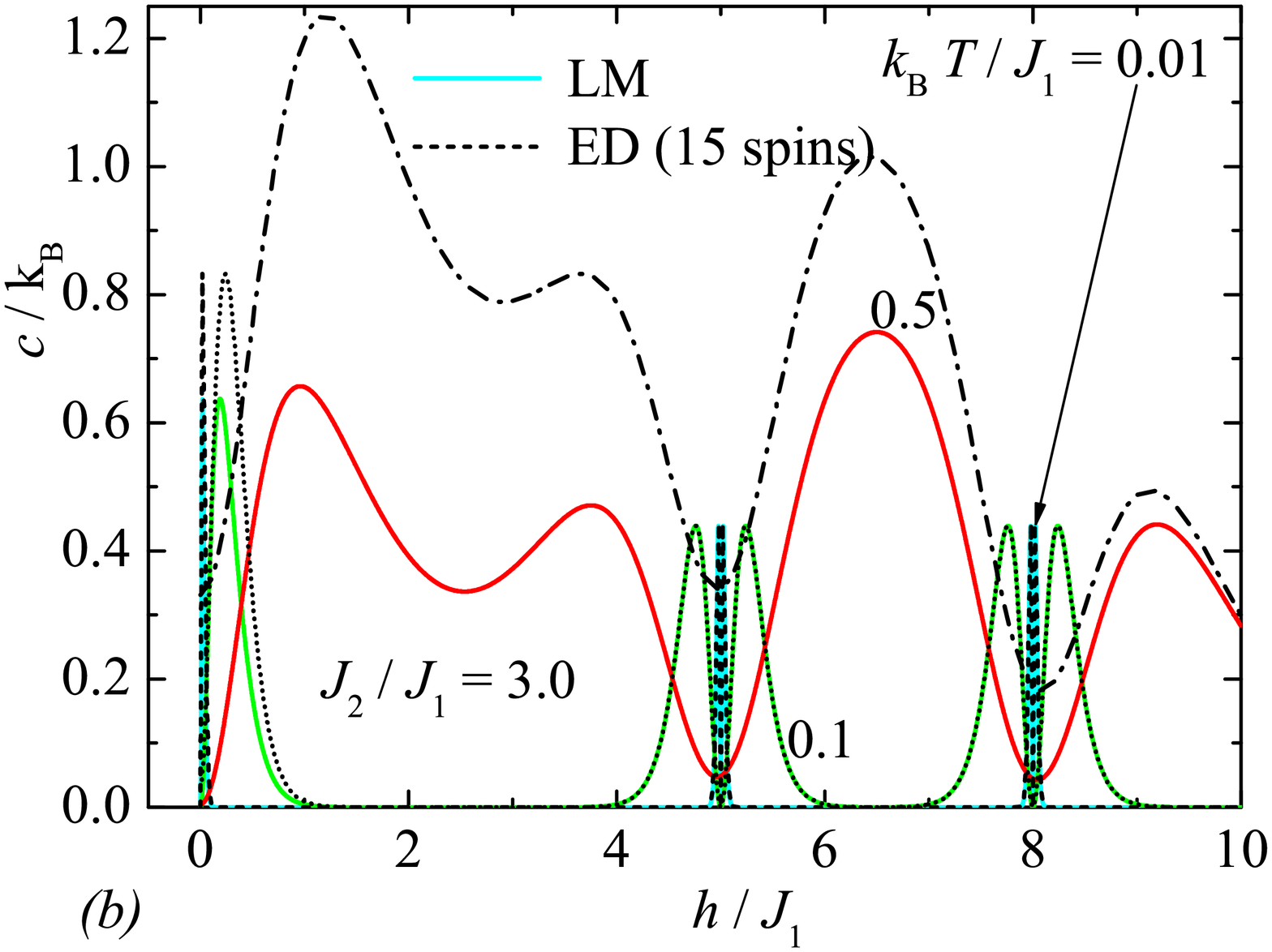}
\includegraphics[width=0.45\textwidth]{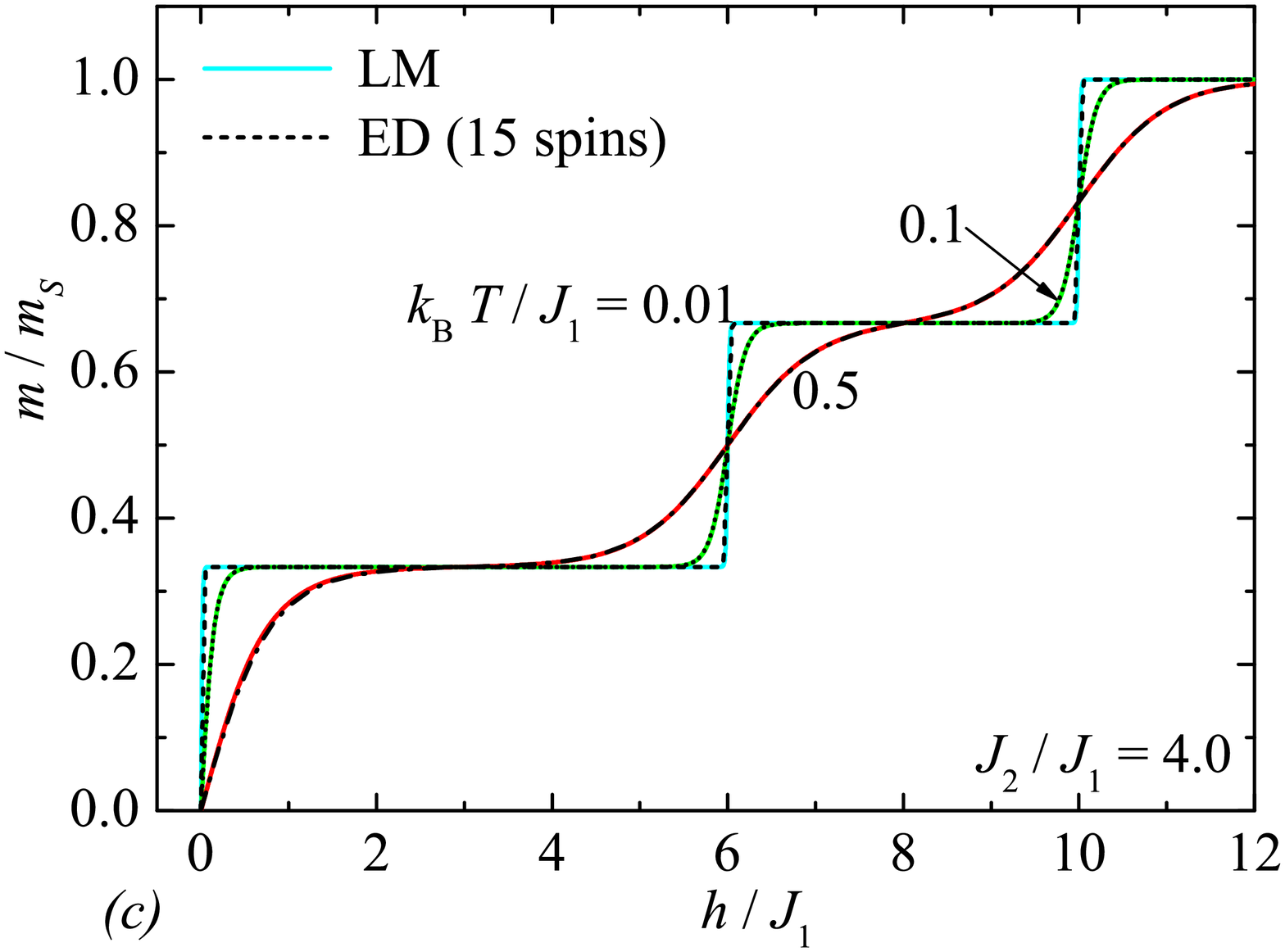}
\hspace{0.5cm}
\includegraphics[width=0.45\textwidth]{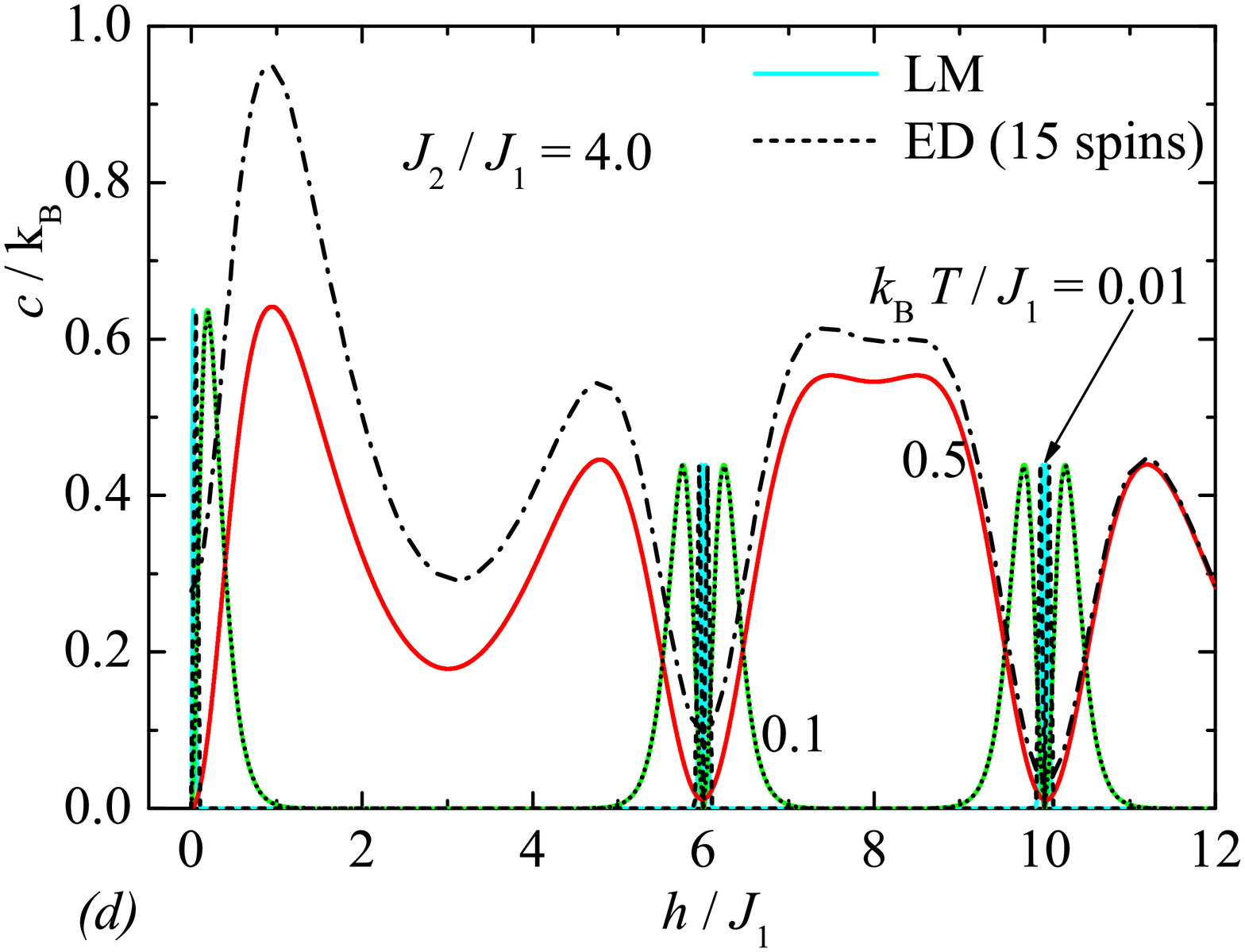}
\end{center}
\vspace{-1.0cm}
\caption{(Color online) The magnetization (left panel) and specific heat (right panel) of the mixed spin-1 and spin-1/2 Heisenberg octahedral chain as a function of the magnetic field for a few different temperatures and the interaction ratio: (a)-(b) $J_2/J_1 = 3$; (c)-(d) $J_2/J_1 = 4$. Solid lines represents the results stemming from the localized-magnon approach, whereas broken lines follow from full ED for a finite-size mixed-spin Heisenberg octahedral chain with $N=3$ unit cells ($15$ spins).}
\label{fig9}
\end{figure*}

In the following part we will take advantage of the localized-magnon approach elaborated in Sect.  \ref{lmt} in order to discuss the most interesting results for a low-temperature  magnetothermodynamics of the mixed spin-1 and spin-1/2 Heisenberg octahedral chain in  a highly frustrated parameter region $J_2/J_1 \geq 3$. The magnetization process of the mixed spin-1 and spin-1/2 Heisenberg octahedral chain is depicted in Fig. \ref{fig9}(a) as a function of the external magnetic field for three different temperatures and the interaction ratio $J_2/J_1=3$ representing a lower limit for applicability of the localized-magnon approach. As one can see from Fig. \ref{fig9}(a), the magnetization curve of the mixed spin-1 and spin-1/2 quantum Heisenberg octahedral chain exhibits two intermediate plateaus at one-third and two-thirds of the saturation magnetization, which correspond to the monomer-tetramer phase (\ref{MT}) and the bound magnon-crystal phase (\ref{LM}) in accordance with the ground-state phase diagram (see Fig. \ref{fig5}). It is obvious that an increase of temperature causes a gradually smoothing of the stepwise magnetization curve. The magnetization data obtained from the localized-magnon approach are in a plausible agreement with the full ED data at low enough temperatures, while both results start to deviate from each other at moderate temperatures $k_{\rm{B}}T/J_1 \approx 0.5$ due to a low-energy excitations presumably towards the hexamer-tetramer ground state. The hexamer-tetramer ground state coexists with the monomer-tetramer ground state at the specific value of the interaction ratio $J_2/J_1=3$, which accordingly represents a lower limit for the usability of the localized-magnon approach. The higher the temperature is, the higher is the deviation between the outcomes of ED and localized-magnon approach. It can be seen from Fig. \ref{fig9}(b) that the specific heat of the mixed spin-1 and spin-1/2 Heisenberg octahedral chain shows even more pronounced differences between the ED data and the respective results gained within the localized-magnon approach. However, both approaches predict the qualitatively same temperature dependence of the specific heat with two separate peaks located in a proximity of critical magnetic fields, at which the magnetization undergoes at zero temperature discontinuous jump. It is evident from Fig. \ref{fig9}(b) that the localized-magnon approach undervalues the specific heat primarily at low magnetic fields, where the excited states related to the hexamer-tetramer state are not accounted for. As a matter of fact, better quantitative agreement can be obtained if one considers higher values of  the interaction ratio $J_2/J_1$, which would fall deeper inside into a phase stability of the monomer-tetramer ground state, e.g. $J_2/J_1=4$. To illustrate the case, the magnetization curve and specific heat of the mixed spin-1 and spin-1/2 of Heisenberg octahedral chain is displayed in Fig. \ref{fig9}(c) and (d) for this particular value of the interaction ratio. As one can see, the magnetization data obtained from the localized-magnon approach are in a perfect agreement with the ED data up to relatively high temperatures $k_{\rm{B}}T/J_1 \lesssim 0.5$ and the same conclusion can be reached for the field dependence of specific heat. It can be understood from Fig. \ref{fig9}(d) that the quantitative discrepancy between the specific-heat data obtained from the localized-magnon approach and full ED gradually diminishes as the interaction ratio $J_1/J$ strengthens. In fact, the localized-magnon approach gives just slightly undervalued specific heat when comparing it with ED data at relatively high temperature ($k_{\rm{B}}T/J_1\simeq 0.5$)
	 due to excited states neglected within the proposed localized-magnon scheme.

\section{Conclusion}
\label{sec:conc}
The present article provides a detailed study of the mixed spin-1 and spin-1/2 Heisenberg octahedral chain by the use of several complementary analytical and numerical methods. The mixed spin-1 and spin-1/2 Heisenberg octahedral chain exhibits a plethora of exotic quantum states with the character of the uniform Haldane phase, the cluster-based Haldane phases, the ferrimagnetic phases of Lieb-Mattis type, the quantum spin liquids and the bound magnon-crystal phases.

The low-temperature magnetothermodynamics was elaborated in the highly frustrated parameter region $J_2/J_1>3$ by the use of localized-magnon approach, which allows to examine a magnetization process and other basic thermodynamic quantities in a full range of magnetic fields from zero up to saturation. This approach is based on a two-component lattice-gas model accounting for the lowest-energy eigenstates being composed of a bound one- and two-magnon states. A comparison between the results obtained from the localized-magnon approach and exact diagonalization has proved a satisfactorily description of low-temperature magnetothermodynamics of the mixed spin-1 and spin-1/2 Heisenberg octahedral chain in a full range of the magnetic field up to moderate temperatures $k_{\rm{B}}T/J \approx 0.5$. 

The most spectacular quantum ground states relate to three  cluster-based Haldane phases, which exhibit a higher-period of a magnetic unit cell due to a spontaneous breaking of the translational symmetry. The cluster-based Haldane phases are constituted from a finite spin cluster in a triplet state (a few connected octahedra), which can be effectively described by the open spin-1 Heisenberg chains with odd number of spins separated from each other by plaquette-singlet state. Whilst two fragmentized cluster-based Haldane phases with the period $p=3$ and 4 are stable only in a relatively narrow parameter region, the hexamer-tetramer phase as another special case with the period $p=2$ is stable in a relatively wide interval of the magnetic fields. It is worthwhile to remark that analogous cluster-based Haldane phase has been recently predicted also for the mineral crystal fedotovite.\cite{fuji18} The cluster-based Haldane phases are subject of current intense interest from the viewpoint of quantum processing of information and quantum computing, because appropriate modification of them could possibly lead to a creation of topologically protected edge states. This represents challenging task for future study. 

\begin{acknowledgments}
This work was supported by the grant of The Ministry of Education, Science, Research and Sport of the Slovak Republic under the contract No. VEGA 1/0043/16 and by the grant of the Slovak Research and Development Agency under the contract No. APVV-16-0186.
\end{acknowledgments}

{\
\appendix
\section{A proof for local conservation of the composite spins $\boldsymbol{\hat{S}}_{\square,j}^2$, $\boldsymbol{\hat{S}}_{24,j}^2$, and $\boldsymbol{\hat{S}}_{35,j}^2$}
\label{appa}
The proof is based on the generalization of the commutation relations for the composite spins $\boldsymbol{\hat{S}}_{\square,j}$, $\boldsymbol{\hat{S}}_{24,j}$, $\boldsymbol{\hat{S}}_{35,j}$. They are the same as for a single spin $\boldsymbol{\hat{S}}_{l,j}$, i.e. 
\begin{eqnarray}
&&[\boldsymbol{\hat{S}}_{\square,j}^x,\boldsymbol{\hat{S}}_{\square,j}^y] = i\boldsymbol{\hat{S}}_{\square,j}^z, \quad
[\boldsymbol{\hat{S}}_{24,j}^x,\boldsymbol{\hat{S}}_{24,j}^y] = i\boldsymbol{\hat{S}}_{24,j}^z, \quad 
\nonumber\\
&&[\boldsymbol{\hat{S}}_{35,j}^x,\boldsymbol{\hat{S}}_{35,j}^y] = i\boldsymbol{\hat{S}}_{35,j}^z, 
\label{composite-cr1}
\end{eqnarray}
and any cyclic permutation of upper indices. As a consequence, the commutation relations for $\boldsymbol{\hat{S}}_{\square,j}^2$, $\boldsymbol{\hat{S}}_{24,j}^2$, $\boldsymbol{\hat{S}}_{35,j}^2$ can be also readily obtained
\begin{eqnarray}
[\boldsymbol{\hat{S}}_{\square,j}^2,\boldsymbol{\hat{S}}_{\square,j}]=
[\boldsymbol{\hat{S}}_{24,j}^2,\boldsymbol{\hat{S}}_{24,j}]=
[\boldsymbol{\hat{S}}_{35,j}^2,\boldsymbol{\hat{S}}_{35,j}]=0.
\label{composite-cr2}
\end{eqnarray}
First, let us rewrite the Hamiltonian (\ref{ham}) in terms of composite spin operators 
$\boldsymbol{\hat{S}}_{24,j}$ and $\boldsymbol{\hat{S}}_{35,j}$
\begin{eqnarray}
\hat{\cal H} \!\!&=&\!\! J_1 \sum_{j=1}^{N}\Bigl[ (\boldsymbol{\hat{S}}_{1,j} + \boldsymbol{\hat{S}}_{1,j+1}) \!\cdot\! 
(\boldsymbol{\hat{S}}_{24, j}+\boldsymbol{\hat{S}}_{35, j})
\nonumber \\  
\!\!&+&\!\! 
J_2\boldsymbol{\hat{S}}_{24, j}\cdot \boldsymbol{\hat{S}}_{35, j}
-h (\hat{S}_{1,j}^{z} +\hat{S}_{24,j}^{z} + \hat{S}_{35,j}^{z}) \Bigr]. 
\label{ham_a}
\end{eqnarray}
It directly follows from Eq.~(\ref{composite-cr2}) that $[\boldsymbol{\hat{S}}_{24,j}^2,\hat{\cal H}]=0$ and $[\boldsymbol{\hat{S}}_{35,j}^2,\hat{\cal H}]=0$ and hence, the composite spin operators $\boldsymbol{\hat{S}}_{24,j}^2$ and $\boldsymbol{\hat{S}}_{35,j}^2$ correspond to conserved quantities with well defined quantum spin numbers.

To prove that the composite spin operator of a square plaquette commutes with the Hamiltonian, i.e. $[\boldsymbol{\hat{S}}_{\square,j}^2,\hat{\cal H}]=0$, it is quite convenient to use the Hamiltonian in the form of Eq.~(\ref{hamlcl}) providing the following result
\begin{eqnarray}
[\boldsymbol{\hat{S}}_{\square,j}^2,\hat{\cal H}]=-\frac{J_2}{2}\left([\boldsymbol{\hat{S}}_{\square,j}^2,\boldsymbol{\hat{S}}_{24,j}^2]
+[\boldsymbol{\hat{S}}_{\square,j}^2,\boldsymbol{\hat{S}}_{35,j}^2]\right).
\label{cr1}
\end{eqnarray}
The commutators given by the last two terms of Eq.~(\ref{cr1}) vanish, because $\boldsymbol{\hat{S}}_{\square,j}^2=\boldsymbol{\hat{S}}_{24,j}^2+\boldsymbol{\hat{S}}_{35,j}^2+2\boldsymbol{\hat{S}}_{24,j}\cdot\boldsymbol{\hat{S}}_{35,j}$ and the composite spin operators $\boldsymbol{\hat{S}}_{\square,j}^2$, $\boldsymbol{\hat{S}}_{24,j}^2$, $\boldsymbol{\hat{S}}_{35,j}^2$ represent according to Eq. (\ref{composite-cr2}) a set of commuting operators. Hence, the commutator on the left-hand-side of Eq.~(\ref{cr1}) also turns to zero and this completes the proof for the local conservation of the total spin of square plaquette represented by the composite spin operator $\boldsymbol{\hat{S}}_{\square,j}^2$. The proof for $\boldsymbol{\hat{S}}_{\square,j}^z$ can be accomplished in an analogous way.

\section{A criterion for existence of cluster-based Haldane phases}
\label{appb}
An existence of the cluster-based Haldane phases is closely connected to a fragmentation of the mixed spin-1 and spin-1/2 Heisenberg octahedral chain into smaller spin clusters with the character of the effective finite-size spin-1 Heisenberg chain under open boundary conditions, which are isolated from each other by plaquette-singlet states. The effective Hamiltonian for the cluster-based Haldane phases developed from the Hamiltonian (\ref{hamlcl}) consequently reads 
\begin{eqnarray}
\hat{\cal{H}}_{1_{2p{-}1}{-}0}\!&=&\! J_1\left[\sum_{l=0}^{\frac{N}{p}{-}1}\sum_{j=1}^{p-1}\left(\boldsymbol{\hat{S}}_{1,lp{+}j}{+}\boldsymbol{\hat{S}}_{1,lp{+}j{+}1} \right) 
{\cdot} \boldsymbol{\hat{S}}_{\square,lp{+}j}\right] \nonumber \\
\!&-&\! \frac{N}{p}J_2(p+1)
\label{cbhf}
\end{eqnarray}
where $p$ is the magnetic period of ground state determining a regular repetition of the plaquette-singlet state. The ground-state energy corresponding to the effective Hamiltonian (\ref{cbhf}) of the period-$p$ cluster-based Haldane phase can be expressed as follows 
\begin{eqnarray}
E_{1_{2p-1}-0} (2N, S_T = \frac{N}{p}) = \frac{N}{p}\left[J_1\varepsilon_{2p-1}^1 - J_2(p+1)\right]. \nonumber \\
\label{cbhfe}
\end{eqnarray}
Here, the symbol $\varepsilon_{2p-1}^1$ denotes the ground-state energy of the spin-1 Heisenberg chain with the odd number of spins $2p-1$ and unit coupling constant, which belongs to the triplet sector with the total spin $S_T=1$. According to the formula (\ref{cbhfe}) the energy of the $(p+1)$-periodic cluster-based Haldane state is smaller than the energy of the $p$-periodic cluster-based Haldane state when the intra-plaquette coupling constant $J_2$ is smaller than the critical value 
\begin{eqnarray}
J_2 (p \to p+1) = J_1 \left[(p + 1) \varepsilon_{2p-1}^1 - p \varepsilon_{2p+1}^1 \right].
\label{J2c}
\end{eqnarray}
A stability condition of the cluster-based Haldane phase with the period $p$, which would be wedged in between the $(p-1)$- and $(p+1)$-periodic cluster-based Haldane phases, is then given by the inequality $J_2(p-1\to p) > J_2(p\to p+1)$ taking the following equivalent form
\begin{eqnarray}
\varepsilon_{2p-1}^1 - 2\varepsilon_{2p+1}^1 + \varepsilon_{2p+3}^1>0.
\label{tarasik}
\end{eqnarray}
If the prerequisite (\ref{tarasik}) would hold for any period $p$, one could prove by induction existence of an infinite series of the cluster-based Haldane states. However, exact diagonalization data for the spin-1 Heisenberg chain with the odd number of spins $2p-1$ imply that the condition (\ref{tarasik}) is met for the period $p=2$, $3$ and $4$ only.}

\end{document}